\documentclass[twocolumn]{aastex631}

\shorttitle{ICL analysis of WHL0137}
\shortauthors{Jim\'enez-Teja et al.}

\begin{document}

\title{RELICS: ICL Analysis of the $z=0.566$ merging cluster WHL J013719.8-08284}

\email{yojite@iaa.es}

\author[0000-0002-6090-2853]{Yolanda Jim\'enez-Teja}
\affiliation{Instituto de Astrof\'isica de Andaluc\'ia--CSIC, Glorieta de la Astronom\'ia s/n, E--18008 Granada, Spain}

\author[0000-0001-7299-8373]{Jose M. V\'ilchez}
\affiliation{Instituto de Astrof\'isica de Andaluc\'ia--CSIC, Glorieta de la Astronom\'ia s/n, E--18008 Granada, Spain}

\author{Renato A. Dupke}
\affiliation{Observat\'orio Nacional, Rua General Jos\'e Cristino, 77 - Bairro Imperial de S\~ao Crist\'ov\~ao, Rio de Janeiro, 20921-400, Brazil}
\affiliation{Department of Physics and Astronomy, University of Alabama, Box 870324, Tuscaloosa, AL 35487}
\affiliation{Department of Astronomy, University of Michigan, 311 West Hall, 1085 South University Ave., Ann Arbor, MI 48109-1107}

\author[0000-0003-2540-7424]{Paulo A. A. Lopes}
\affiliation{Observat\'orio do Valongo, Universidade Federal do Rio de Janeiro, Ladeira do Pedro Ant\^onio 43, Rio de Janeiro RJ 20080-090, Brazil}

\author[0000-0002-8742-0643]{N\'icolas O. L. de Oliveira}
\affiliation{Observat\'orio Nacional, Rua General Jos\'e Cristino, 77 - Bairro Imperial de S\~ao Crist\'ov\~ao, Rio de Janeiro, 20921-400, Brazil}

\author[0000-0001-7410-7669]{Dan Coe}
\affiliation{AURA for the European Space Agency (ESA), Space Telescope Science Institute, 3700 San Martin Drive, Baltimore, MD 21218, USA}

\begin{abstract}
We present a pilot study of the intracluster light (ICL) in massive clusters using imaging of the $z=0.566$ cluster of galaxies WHL J013719.8-08284 observed by the RELICS project with the HST. We measure the ICL fraction in four optical ACS/WFC filters (F435W, F475W, F606W, and F814W) and five infrared WFC3/IR bands (F105W, F110W, F125W, F140W, and F160W). The ICL maps are calculated using the free of a priori assumptions algorithm CICLE, and the cluster membership is estimated from photometric properties. We find optical ICL fractions that range between $\sim$6\% and 19\% in nice agreement with the values found in previous works for merging clusters. We also observe an ICL fraction excess between 3800 \AA~ and 4800 \AA, previously identified as a signature of merging clusters at $0.18<z<0.55$. This excess suggests the presence of an enhanced population of young/low-metallicity stars in the ICL. All indicators thus point to WHL J013719.8-08284 as a disturbed cluster with a significant amount of recently injected stars, bluer than the average stars hosted by the cluster members and likely stripped out from infalling galaxies during the current merging event. Infrared ICL fractions are $\sim$50\% higher than the optical ones, which could be signature of an older and/or higher-metallicity ICL population that can be associated with the build-up of the BCG, the passive evolution of young stars, previously injected, or preprocessing in infalling groups. Finally, investigating the photometry of the cluster members, we tentatively conclude that WHL J013719.8-08284 fulfills the expected conditions for a fossil system progenitor.
\end{abstract}


%

\section{Introduction}\label{sect_intro}

Cluster of galaxies encode valuable information of the formation and evolution of the Universe. They are the latest and most massive structures to form, according to the standard cosmological model Lambda-Cold Dark Matter \citep[$\Lambda$CDM, ][]{bond1991,lacey1993}. They are built hierarchically, accreting matter along large-scale filaments of the Universe, mainly cannibalizing nearby galaxies and merging with groups and other clusters. In this scenario, the clusters' formation and evolution is plagued with violent processes that strip out stars from the infalling systems, sometimes having their galaxies completely destroyed during the merger. As a result, many of these stars become part of the brightest cluster galaxy (BCG) and other massive galaxies of the main cluster, while others end up floating freely in the intracluster space, completely disconnected from their progenitor galaxies and just gravitationally bound to the cluster's potential. These stars form the so-called intracluster light (ICL), an extended, diffuse, luminous component that surrounds the cluster galaxies and permeates the space between them. \\

The ICL has attracted an increasing amount of attention in the last years, although its study has always been limited by both technical and analytical reasons. The ICL is characterized by a very low surface brightness, typically $\mu_V>26.5$ mag arcsec$^{-2}$ \cite[e.g., ][]{rudick2006}, what makes its detection very complex, specially from ground-based telescopes, needing long exposure times and excellent seeing. Early works that first detected and measured the ICL observed it as faint as 1\% of the sky brightness \citep{feldmeier2002,rudick2006}. Indeed, the sky background plays a crucial role in the analysis of the ICL. In general, standard pipelines of background estimation can easily misidentify ICL with sky, so that many authors develop their own methods to either refine the standard background \citep[e.g. ][]{zibetti2005,krick2007,jimenez-teja2016, montes2018, morishita2017,furnell2021} or estimate it from scratch, using the pre-processed, pre-coadded, individual exposures \citep{gonzalez2005,zhang2019,montes2021}.\\

Another crucial issue is the identification of the stars that belong to the ICL, as opposed to those confined in the galaxies haloes. Traditional techniques define this transition happening at a fixed limit in the radial distance from the galaxies \citep[tipically, $R>50$ kpc, e.g., ][]{montes2014,montes2018,kluge2021,furnell2021}, or the integrated brightness \citep{kluge2021}, or the surface brightness \citep{burke2012,montes2014,montes2018,kluge2021,furnell2021}, or the flux emitted over the background level as provided by SExtractor \citep{bertin1996} segmentation map \citep{feldmeier2002,zibetti2005,krick2006}. Other classical methods include fitting the galaxies using single or composite analytical profiles, such as Gaussian, Sersic, de Vaucouleurs, and exponential profiles \citep{kluge2021}. A further complication is disentangling the ICL from the brightest cluster galaxy (BCG), which usually has an extended halo that smoothly submerges in the ICL. Moreover, as the ICL dynamics is governed by the cluster potential, these stars are mostly concentrated around the cluster core, close to the BCG. The projected spatial coincidence of these two luminous surfaces (ICL and BCG) in the plane of the cluster, makes it very complex to truly know where the transition occurs \citep{rudick2011}. Many works do not attempt to separate these two components, but they consider the BCG+ICL system as an unity \citep{lin2004,zibetti2005,toledo2011}. Other authors try to fit the BCG (and the ICL) with single (double), very extended analytical profiles \citep{gonzalez2005,zibetti2005}, whereas others apply the same criteria to the BCG and the rest of the cluster galaxies \citep{krick2006}. The best approach would be disentangling the BCG from the ICL by their kinematic properties \citep{coccato2011, toledo2011}. It is important to note that, several authors showed that the transitions found by the previously described criteria do not match those obtained from the kinematics. For example, \cite{rudick2011} showed with simulations that the surface brightness method systematically yields lower ICL fractions (ratio between the ICL and the total cluster luminosity) than those found by binding energy and/or a kinematic analysis, loosing between 30\% to 80\% of the ICL stars. Analogously, \cite{cui2014} used simulated data to disentangle the BCG and the ICL according to the different velocity distributions of their stellar particles, and proved that the transition between these two components occurred at much smaller radii than those provided by surface brightness cuts. In fact, the difference in the ICL fractions provided by the two methods was a factor of three or four, on average. Also using mock clusters, \cite{cooper2015} traced back dark matter particles to disentangle in situ stars (formed directly from the cluster cooling flow) from accreted stellar material (stripped from galaxies other than the BCG) and found that this was not equivalent to the separation provided by fitting with two Sersic functions. Instead, this modelling seemed to trace the transition between relaxed and unrelaxed accreted components, which is not equivalent to the dynamical definition of ICL. This latter approach was further investigated by \cite{kluge2021} with real data, concluding that the double Sersic decomposition is likely to be nonphysical.\\

Another typical uncertainty in photometric ICL analyses is the disparity of results among the different traditional methods and within a certain technique (highly dependent on the value of the galaxy-ICL threshold), which leads to conclusions that are often inconsistent and, sometimes, even contradictory. Several authors indeed quantified the impact of the chosen threshold-modeled parameters and their values on the final results; for instance, \citep{rudick2011} found that the ICL fraction can differ up to a factor of two depending on the value of the surface brightness limit, using simulations, which was later confirmed observationally by \cite{burke2012}. More recently, \cite{kluge2021} confirmed that this discrepancy can rise even more comparing all classical methods applied on real data. In spite of these findings, the simplicity of the traditional techniques makes them easily replicable and comparable, which are clear advantages. Indeed, some authors advocate the use of an universal criteria to define the ICL that, even yielding results that may not be quantitatively reliable, can provide qualitative information about the ICL \citep{pillepich2018}, such as trends or correlations with physical properties of the cluster.  \\

There are also new and more sophisticated techniques that tackle the problem of the ICL estimation from a more sophisticated mathematical angle. Among those, we can remark the wavelet approach \citep{darocha2008,adami2005,guennou2012,adami2013}, the fitting and averaging technique developed \citep{morishita2017}, and the curvature-based algorithm CICLE \citep{jimenez-teja2016,jimenez-teja2018,jimenez-teja2019}. These algorithms are often tested against simulated data, probing their accuracy and efficiency, which are often superior to those of the traditional techniques. However, their implementation is complex and the physical interpretation of their parameters (which are frequently purely mathematical) is not straightforward, which are caveats that prevent these methods from being widely used. \\

 In the last decade, several projects have focused on collecting the deepest images of clusters and groups of galaxies. Among the most recent examples we can mention the Frontier Fields initiative \citep{lotz2017}, the Hyper Suprime-Cam Subaru Strategic Program \citep{aihara2018a,aihara2018b,aihara2019}, the Fornax Deep Survey \citep{iodice2016}, or the upcoming Vera C. Rubin Observatory \cite[previously known as the Large Synoptic Survey Telescope, ][]{ivezic2008}. In addition, the technical development offers a new and barely explored perspective: a panchromatic, more detailed view of the Universe through the use of narrow-band filter systems, as those from the J-PLUS and J-PAS surveys \citep{cenarro2019,benitez2014,bonoli2021}. In this work we intend to show how the combination of these two characteristics, high-quality data and information at different wavelengths, can provide strong clues on the clusters' formation and evolution through the analysis of the ICL. It is known that the ICL presents different morphology, substructures, and ICL fractions depending on the wavelength used \citep{tang2018}. Whereas the ICL fraction measured in single bands can be useful for establishing qualitative relations (e. g., evolution with redshift or halo mass), a panchromatic view of the ICL can give insights on its stellar populations or the dynamical stage of the cluster \citep[e.g.][]{jimenez-teja2019}. With this aim, we use the recently acquired data from the Reionization Lensing Cluster Survey \citep[RELICS, ][]{coe2019}, a Hubble Space Telescope (HST) Treasure program that has observed 41 very massive, strong-lensing clusters in 3 optical (ACS) and 4 infrared (WFC3) bands. In this work we present the potential of the RELICS data for ICL purposes, focused on the cluster WHL J013719.8-08284 (WHL0137 hereafter). Future work will apply the CICLE algorithm, which is free of {\it a priori} assumptions, extensively to the whole RELICS sample with two main aims: 1) analyze the optical and infrared ICL fractions to extract information about the stellar populations of the diffuse light, and 2) infer the dynamical stage of the clusters.   \\
 
 \cite{krick2007} linked some properties of the ICL with the dynamical state of the clusters, using a sample of 10 systems spanning the redshift interval $0.05<z<0.3$. They considered several indicators of the dynamical state of the clusters: a) the magnitude gap between the first and third-ranked galaxies ($\Delta m_{1,3}$; the larger the gap, the higher the relaxation), b) the density (the higher the density, the higher the number of interactions), and c) the X-ray morphology. They found that the ICL flux was correlated with density, and moderately anticorrelated with $\Delta m_{1,3}$. Similarly, systems with more symmetric hot gas distribution had both less ICL and significantly steeper profiles. \cite{pierini2008} also found different properties in the ICL for merging and relaxed clusters. In \cite{jimenez-teja2018} we quantified for the first time the impact that a system's dynamical stage could have on the ICL fractions calculated at different wavelengths. We showed that the optical ICL fractions in relaxed clusters are almost independent of the wavelengths and, on average, 2.3 times lower than those of merging clusters, indicative that many of the stars thrown to the ICL are connected to the merging event. Moreover, we showed that the merger imprint would be more remarkable in the ICL fractions measured between 3800 \AA --4800 \AA~ in the rest-frame. The sample of clusters considered spanned the redshift interval $0.18<z<0.55$. As WHL0137 is in the upper limit of this interval ($z=0.566$), we conduct a similar analysis to that described in \cite{jimenez-teja2018}, and we will anchor our new results to those.  \\
 
This paper is organized as follows: we first describe the main properties of WHL0137, the data collected and the main features of the reduction process in Sect. \ref{sect_data}. The methodology applied to estimate the ICL fractions is explained in Sect. \ref{sect_methodology}, splitting the main steps in three: calculation of the ICL maps with CICLE, refinement of the sky background, and cluster membership from photometric information. Sect. \ref{sect_results} offers a description of the results, along with an extensive discussion on them. Finally, conclusions are wrapped up in Sect. \ref{sect_conclusions}. Throughout this paper we will assume a standard $\Lambda$CDM cosmology with $H_0=70$ km s$^{-1}$ Mpc$^{-1}$, $\Omega_m=0.3$, and $\Omega_{\Lambda}=0.7$. All the magnitudes are referred to the AB system.

\section{Data}\label{sect_data}

RELICS \citep{coe2019} is a 188-orbit Hubble Treasure Program that observed 41 clusters of galaxies spanning the redshift interval $0.182\leq z\leq0.972$ with the Advanced Camera for Surveys (ACS) and the Wide Field Camera 3 (WFC3) on board the Hubble Space Telescope (HST). Clusters were selected according to two main criteria: 1) 21 of them are among the most massive clusters identified by Planck \citep{planck2016}, and 2) the remaining 20 are identified as powerful strong lenses. The main goals of the project can be summarized in the search for high-redshift galaxies and supernovae, studying the strong lensing effect on these clusters, improving the accuracy of the mass scaling relations and imposing constraints on dark matter cosmological parameters.\\

Observations were carried out in the optical and the near infrared (IR), and joint with pre-existing, archival data. Each one of the 41 clusters was observed to 5-orbit depth. None of the RELICS clusters had existing IR HST data, so two orbits were divided among the four WFC3/IR filters F105W, F125W, F140W, and F160W. The requirement was to reach single-orbit depth for each of the three ACS filters F435W, F606W, and F814W, so the three remaining orbits (minus the archival data) were split according to this. For the particular case of WHL0137, additional HST images were observed, as it is described below. \\

WHL0137 (R.A. = 1$^{\rm h}$37$^{\rm m}$25$^{\rm s}$.0, Dec = -8$^{\circ}$27'25 0'' [J2000.0]), is named in the literature as WHL J013719.8-082841, WHL J24.3324-8.477, PSZ1 G155.25-68.42, PSZ2 G155.27-68.42, and [RRB2014] RM J013725.0-082722.7. It was discovered by \cite{wen2012} based on photometric redshifts from the Sloan Digital Sky Survey III (SDSS), detecting the luminous galaxies and appyling a friends-of-friends algorithm to identify the clusters members. WHL0137 is an intermediate-redshift cluster with $z=0.566$ \citep{wen2015}, as determined from SDSS DR12 spectroscopic data \citep{alam2015}. WHL0137 is ranked as the 31st most massive cluster identified in the Planck PSZ2 catalogue, with $M_{500}\sim 10.99\times 10^{14}\,M_{\sun}$ as determined by this collaboration \citep{ade2016}. RELICS provided the first HST observations of WHL0137, which led to the discovery of 15''-long arc at redshift $z\sim6.2$ \citep{salmon2020}. Being the longest known arc at $z>6$ (Welch et al. in prep.), the RELICS collaboration obtained additional HST imaging in the F475W, F814W, and F110W bands (PI Salmon; GO 15842). Also, WHL0137 will be observed with the NIRCam camera onboard the James Webb Space Telescope (PI Coe; GO 2282), in the near IR filters F090W, F115W, F150W, F200W, F277W, F356W, F410W, and F444W. Final filters and depths used in this work are listed in Table \ref{table_data}.\\

\begin{table*}
\caption{WHL0137 data}\label{table_data}
\centering
\begin{tabular}{ccccc}
   Camera/channel & Filter & \# HST orbits & Depth & Surface brightness limit\\
   & & [orbits] & [AB mag] & [mag arcsec$^{-2}$]\\
\hline
ACS/WFC & F435W & 1 & 27.2 & $28.8\pm 0.4$\\ 
ACS/WFC & F475W & 2 & 27.9 & $29.7\pm 0.5$\\
ACS/WFC & F606W & 1 & 27.6 & $29.7\pm 0.5$\\
ACS/WFC & F814W & 6 & 28.0 & $30.1\pm 0.7$\\
WFC3/IR & F105W & 0.6 & 26.7 & $29.3\pm 0.6$\\
WFC3/IR & F110W & 2 & 27.7 & $30.4\pm 0.9$\\
WFC3/IR & F125W & 0.3 & 26.0 & $28.7\pm 0.6$\\
WFC3/IR & F140W & 0.3 & 26.2 & $28.8\pm 0.6$\\
WFC3/IR & F160W & 0.8 & 26.5 & $29.2\pm 0.5$\\
\hline
\end{tabular}
\end{table*}

Images were reduced by the RELICS collaboration using the standard calibration pipelines: CALACS\footnote{http://www.stsci.edu/hst/acs/performance/calacs\_cte/calacs\_cte.html} for ACS data and CALWF3\footnote{http://www.stsci.edu/hst/wfc3/pipeline/wfc3\_pipeline} for WFC3 images. Calibration steps include bias, dark, flat-field, bias-striping, crosstalk, and CTE corrections. Additionally, persistence masks and improved bad pixel masks have been applied to the IR images. Individual exposures were later aligned following the same prescription as for the Frontier Fields data \citep{lotz2017}. As a result, fully reduced, coadded images are made publicly available by the RELICS collaboration in two different resolutions: 0.06 and 0.03 arcsec/pixel. In this work, we used the 0.06 arcsec/pixel images, since we are not interested in the small-scale features of the ICL but, instead, in increasing the signal-to-noise of the low surface brightness structures.\\

\section{Methodology}\label{sect_methodology}

The ICL fraction, defined as the ratio between the ICL and the total luminosity of the cluster, provides valuable information on the cluster's dynamical stage and the properties of the stellar populations of the ICL in comparison to those hosted by the cluster galaxies. In addition, as the ICL fraction is a single parameter, its interpretation is more simple than a two-dimensional ICL map. To calculate the ICL fraction in a certain filter we need: 1) an ICL map free of galactic light and with the background removed; and 2) a map containing just the galaxies identified as cluster members, thus excluding background and foreground sources (interlopers). We used an algorithm based on galaxy modelling to remove the galactic luminous contribution from our images, a simple algorithm to refine the background subtraction, and a machine learning (ML) algorithm to identify the cluster members using photometric data.\\

\subsection{CICLE}\label{sect_cicle}

The CHEFs IntraCluster Light Estimator \citep[CICLE, ][]{jimenez-teja2012} is an unbiased, accurate, free-of-assumptions algorithm developed to build ICL and total cluster light maps, to later estimate ICL fractions. CICLE's main advantage is the use of models to remove the galactic light instead of assuming certain properties of the ICL. These models are not rigid nor assume a certain shape, but are built from a set of analytical functions that constitute a mathematical basis: the so-called Chebyshev-Fourier functions \citep[CHEFs, ][]{jimenez-teja2012}. As a result, the projected luminous distribution of each galaxy in the image is fitted by a weighted combination of CHEF functions. The precision of the model does not depend on the morphology of the galaxy, neither requires any kind of symmetry. The extension of the model, that is, the radius up to which the galaxy wings are fitted, just depends on the quality of the data: the algorithm alone stops when the wings submerge into the sky noise, or when the wings converge asymptotically. When a galaxy is embedded into ICL, its projected light appears as a bump raising over the ICL level. If, in projection, the galaxy is located away from the BCG, it is straightforward for CHEFs to fit this ``protuberance''. However, when the galaxy is very close to the cluster center and has an extended halo, as normally happens with the BCG, its wings smoothly merge into the ICL and it is very complex to know where the CHEF model should end. For these cases, CICLE incorporates a parameter that finds this transition: a curvature parameter. The curvature is a property of each point in a surface, defined as the difference in the slope between that point and the surroundings. Given that the ICL is a component of the clusters formed by stars with a different kinematic behaviour than those locked up in the galaxies and not bound by their gravitational potentials \citep{arnaboldi2004,gerhard2005,toledo2011,longobardi2018a,longobardi2018b}, we can assume that its surface distribution has a different slope than that of the BCG extended halo. Indeed, this assumption is rather natural, since numerous observational and theoretical studies have shown that the ICL and the BCG have distinct formation pathways \citep{gonzalez2005,mihos2005,burke2012,montes2014,edwards2016,morishita2017,montes2018}. Although they are intimately linked before $z\sim 1$, the ICL has additional channels of injection of stars, specially at $z<0.5$, connected to the assembly history of the cluster as a whole. As a consequence, we expect stellar components with different properties for the ICL and the BCG (color, age, metallicity,...), which implies that these two components must certainly be photometrically distinguishable.\\

Once all the galaxies in the image are fitted, the models are removed to obtain an image just containing ICL and background. Please note that, in the same way the CHEF models are removed, they can be reinserted to build an image with the cluster members alone, given that a catalogue with spectroscopically or photometrically identified members is provided (see Section \ref{phot_memb_ML}).\\

\subsection{Background estimation}\label{sect_bg}

Our HST images were reduced by the RELICS team using the state-of-the-art, optimized pipelines for the ACS and WFC3 cameras and coadded following the same recipe as for the Frontier Fields data. However, the background estimation provided by these pipelines is often not suitable for low-surface-brightness, extended sources as the ICL. In fact, it is usual to find regions surrounding bright objects (as the BCG), where the background is oversubtracted \citep{furnell2021,montes2021}. Thus, a refinement of the standard background is mandatory to calculate more precise ICL fractions. Both the ACS/WFC and WFC3/IR channels are characterized by a small field of view (202"x202" and 123"x137", respectively). As our cluster WHL0137 presents an ICL emission that extends for a significant area of the field of view (especially in the IR filters), few pixels (if any) can be identified as blank or unpolluted by either ICL, galatic light, or border effects. The ideal approach would be using close, parallel fields observed under the same conditions (camera, channel, filter, exposure time, epoch,...) to estimate the background, as in \cite{jimenez-teja2016}. However, these data are rarely available so we decided to adopt the procedure described by \cite{morishita2017}. We approximated the sky by a constant value, estimated from histograms built using the blank pixels in the image. \\

\begin{figure}[]
\centering
\includegraphics[width=9cm]{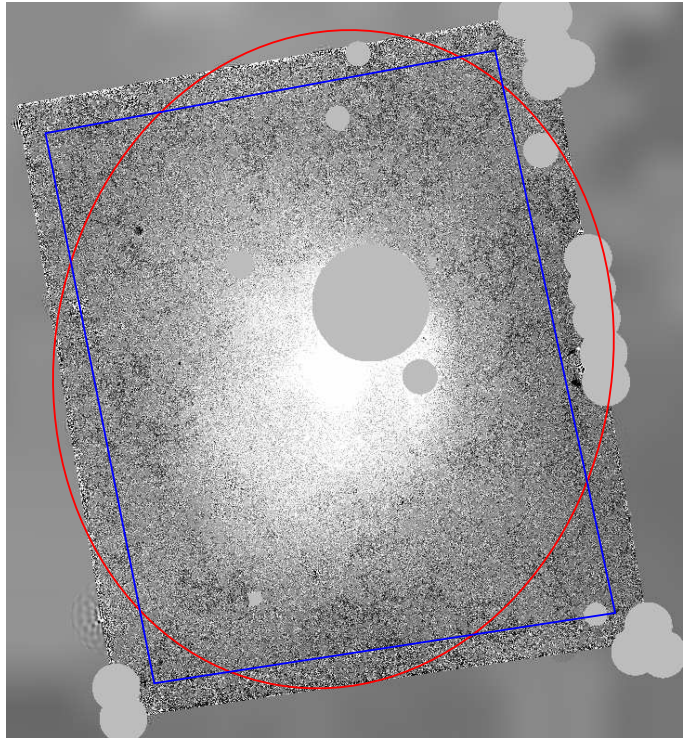}
\caption{Definition of the ICL-free areas over the ICL map calculated by CICLE for the F110W filter. We took as blank areas the regions between the red ellipse and the blue polygon, excluding those pixels that are associated to any source according to the SExtractor segmentation map. Gray circles indicate masked areas due to the presence of stars or very bright sources, before applying CICLE.}
\label{fig_F110Wbg}
\end{figure}

We looked for blank, ICL-free areas close to the corners of the images. However, we note here that, given the small field of view of both cameras and the fact that the ICL in WHL0137 is very extended (especially, in the IR filters), every pixel in the image has some level of contamination by ICL. We thus accept that our refined background will be likely oversubtracted, in the same way that all the algorithms to estimate the background also tend to overestimate it \citep{borlaff2019}. For each filter, we plotted an ellipse centered in the BCG and aligned with the ICL (i.e., with its semimajor axis matching the position angle of the ICL), large enough to contain most of the diffuse light. We also discarded the pixels from the borders since they have some residual, artificial light, to avoid any bias. We used the pixels that were between these two regions and did not belong to any source (according to SExtractor segmentation map) to estimate the background. We remark here that the original segmentation map provided by SExtractor was previously enlarged using a maximum filter of 10x10 pixels to make the area corresponding to each source in the image larger and to ensure that we were not taking the wings of the galaxies as part of the background. Despite we decreased the area of the blank regions with this step, we checked that we still had enough pixels to estimate the background properly. For the sake of illustration, we show in Fig. \ref{fig_F110Wbg} the ICL map in the filter F110W (as calculated by CICLE) along with the inner ellipse and the outer polygon that define the blank regions. Later, we made a histogram with these blank pixels for each one of the filters, using the original HST images. We emphasize that the ICL map (with the contrast enhanced by passing a wavelet filter) was only used to “see” the approximate extension of the ICL and plot the ellipse, but the measurements are done in the original, untouched images. Then, we fitted a Gaussian to each histogram, and took its mean ($\mu$) and standard deviation ($\sigma$) as those of the background. \\

\begin{figure*}[]
\centering
\includegraphics[scale=0.30]{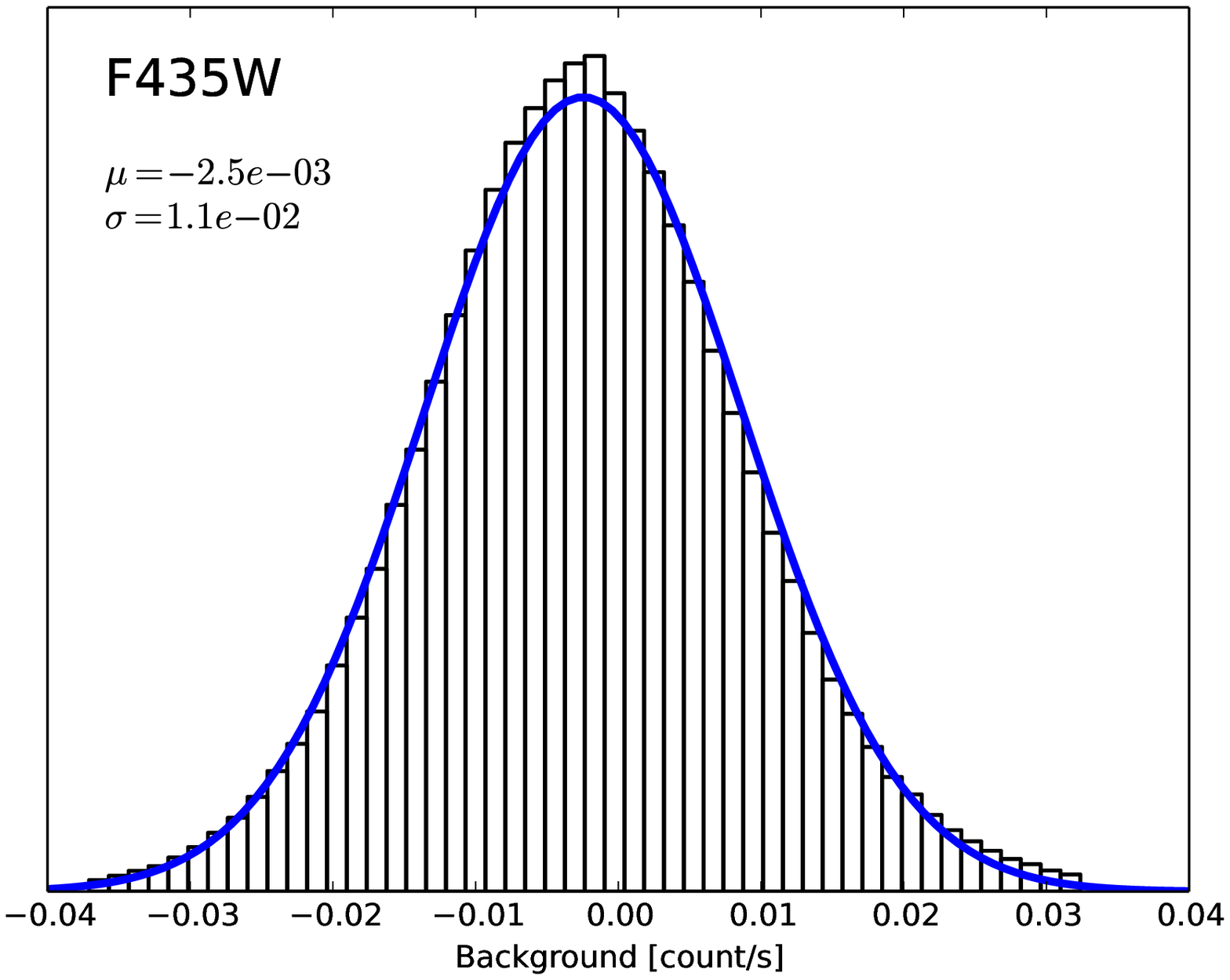}\includegraphics[scale=0.30]{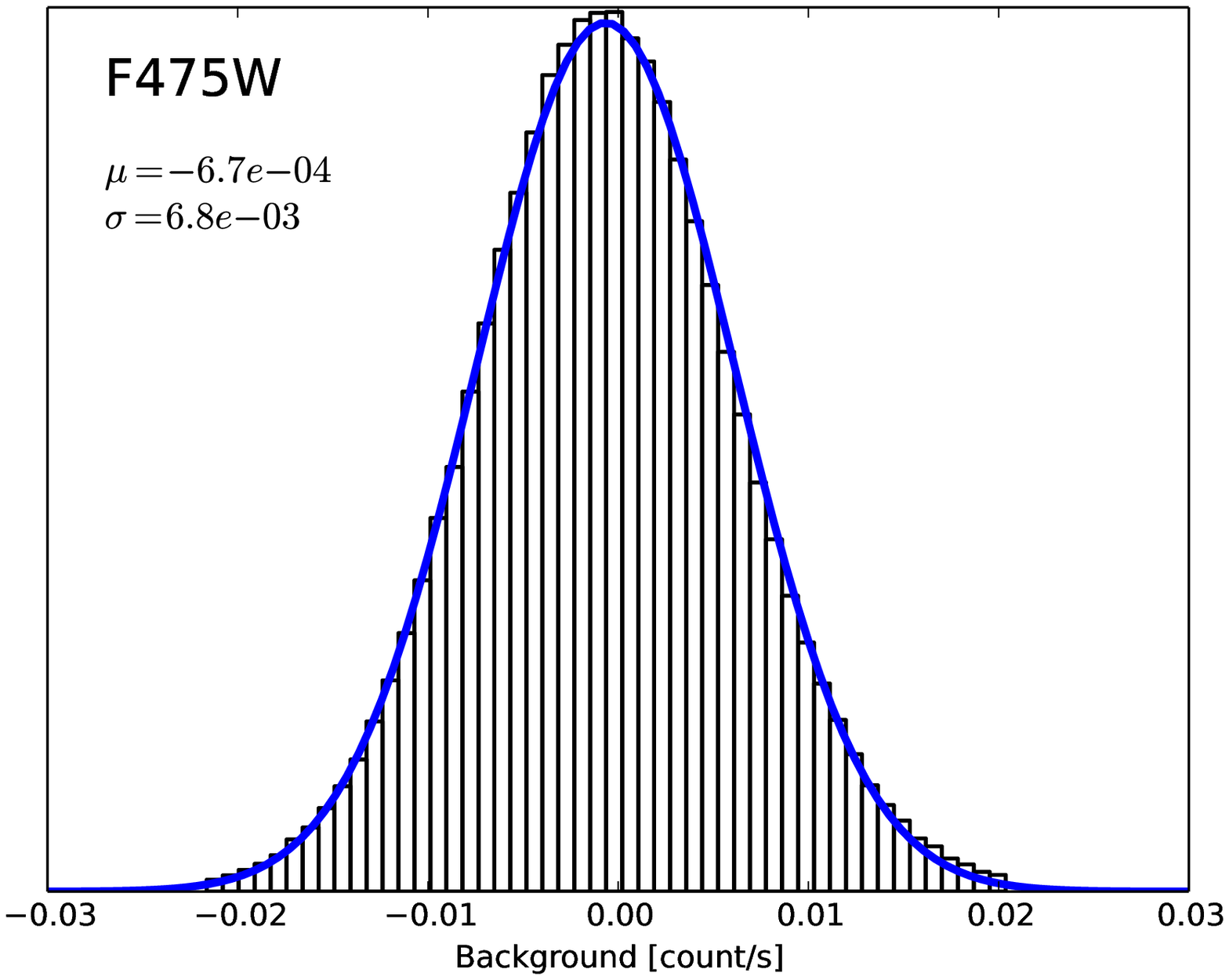}\includegraphics[scale=0.30]{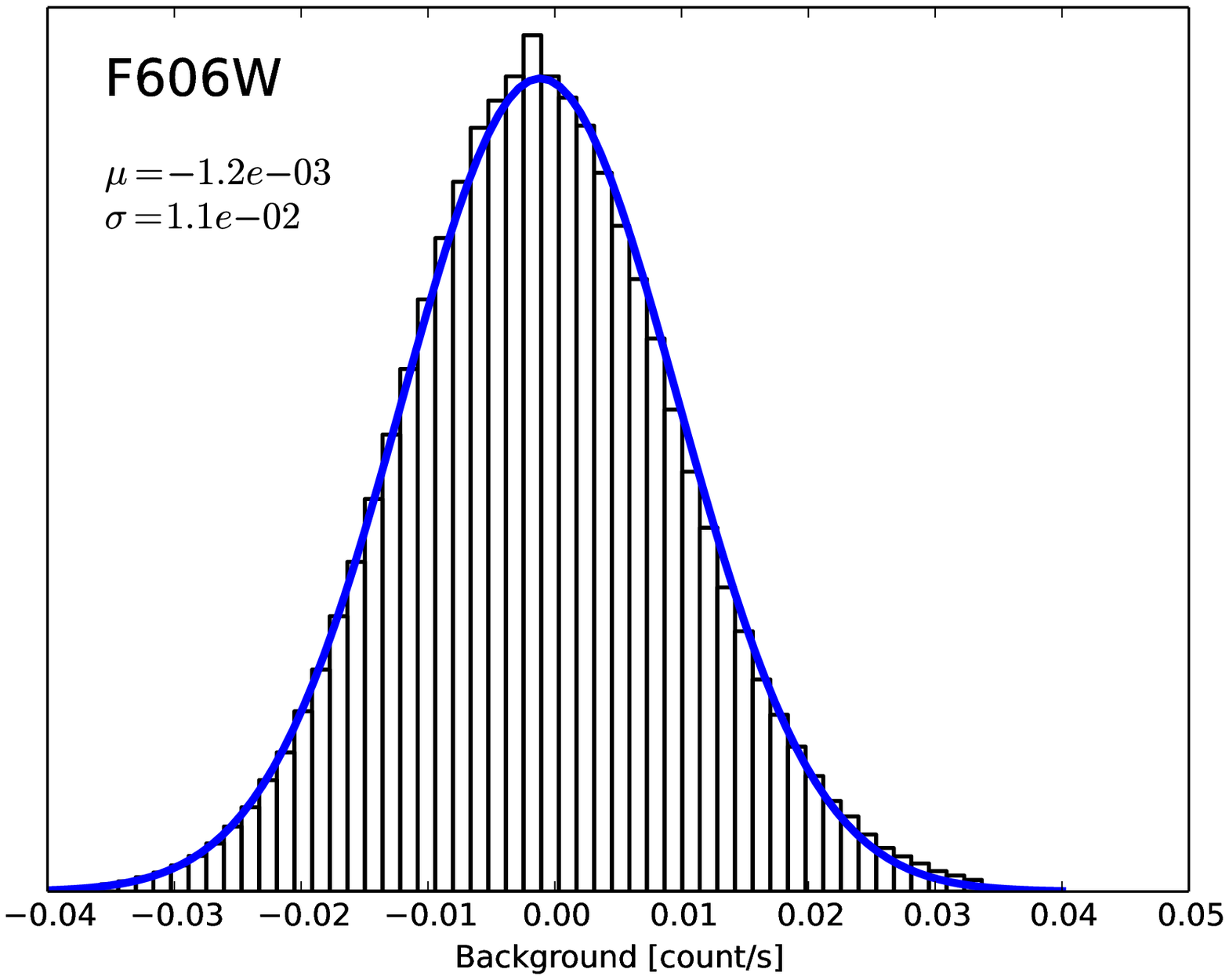}\\
\includegraphics[scale=0.30]{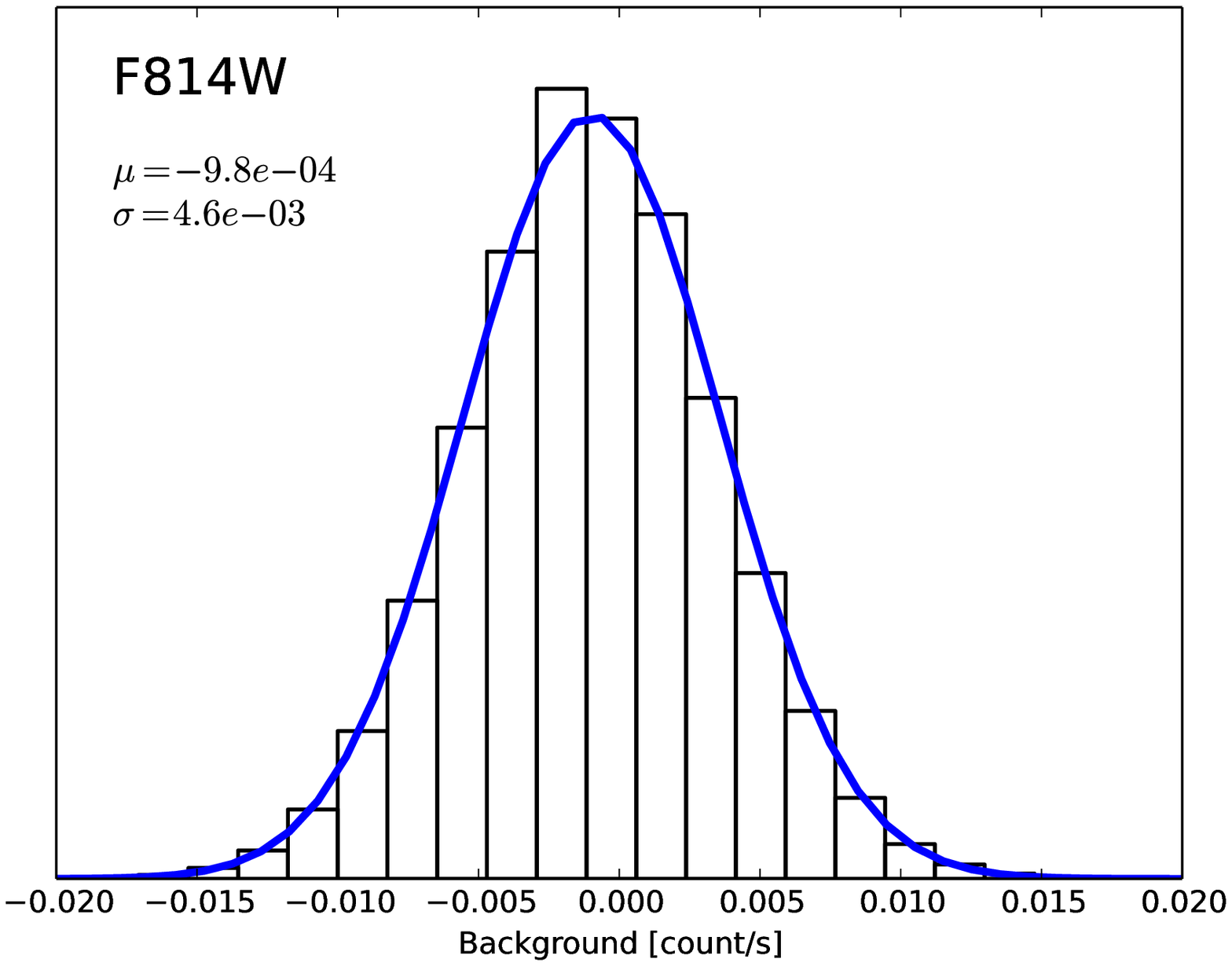}\includegraphics[scale=0.30]{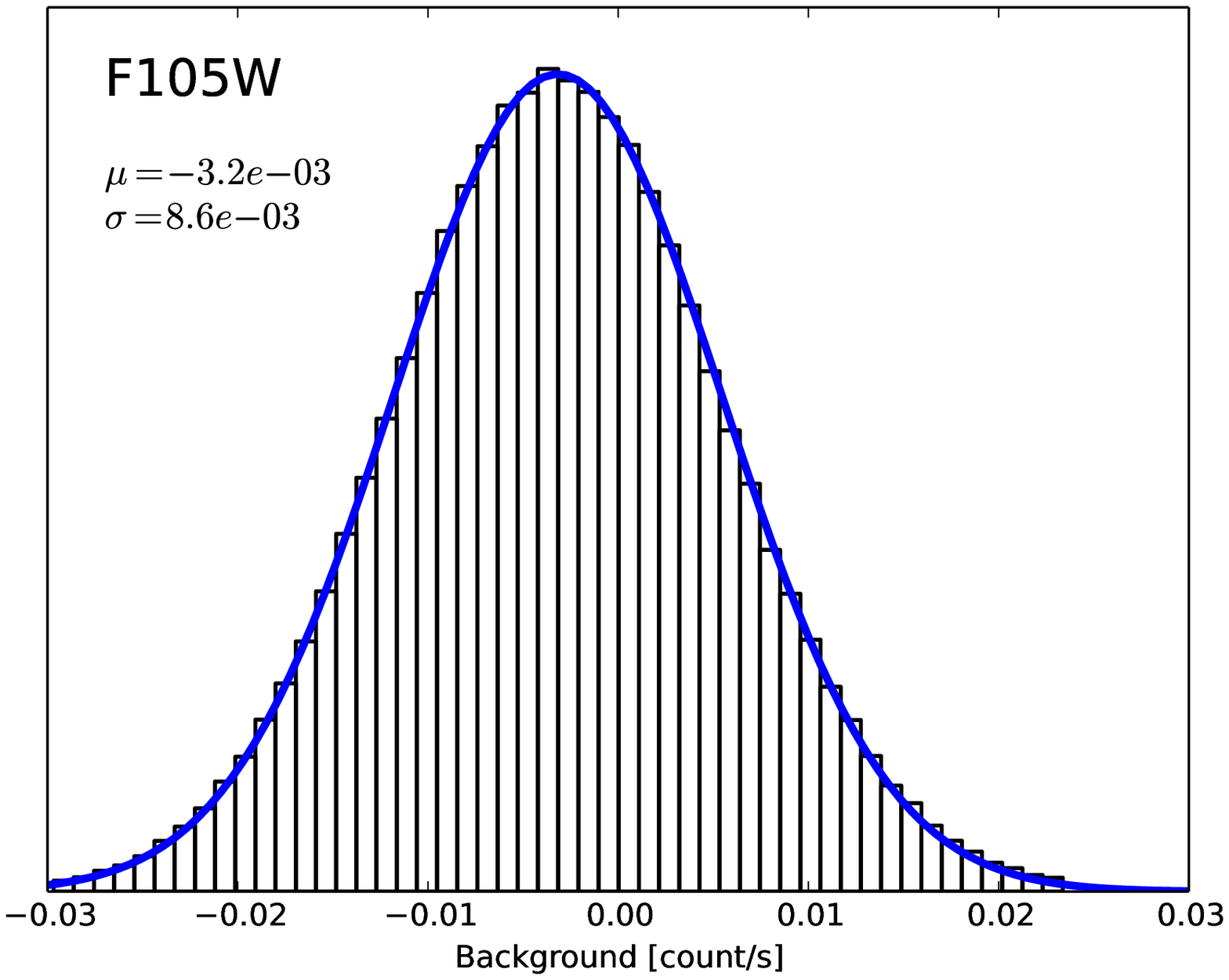}\includegraphics[scale=0.30]{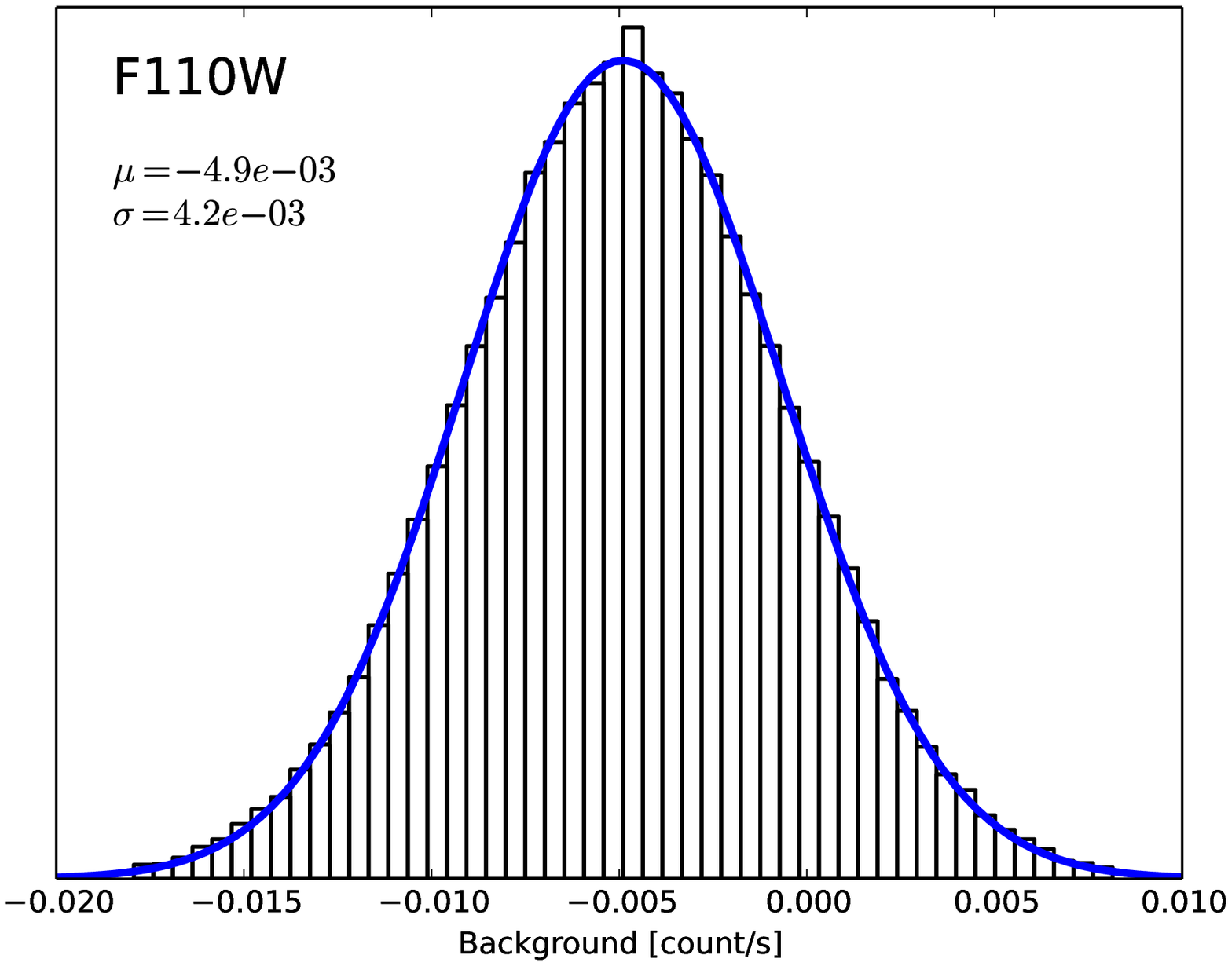}\\
\includegraphics[scale=0.30]{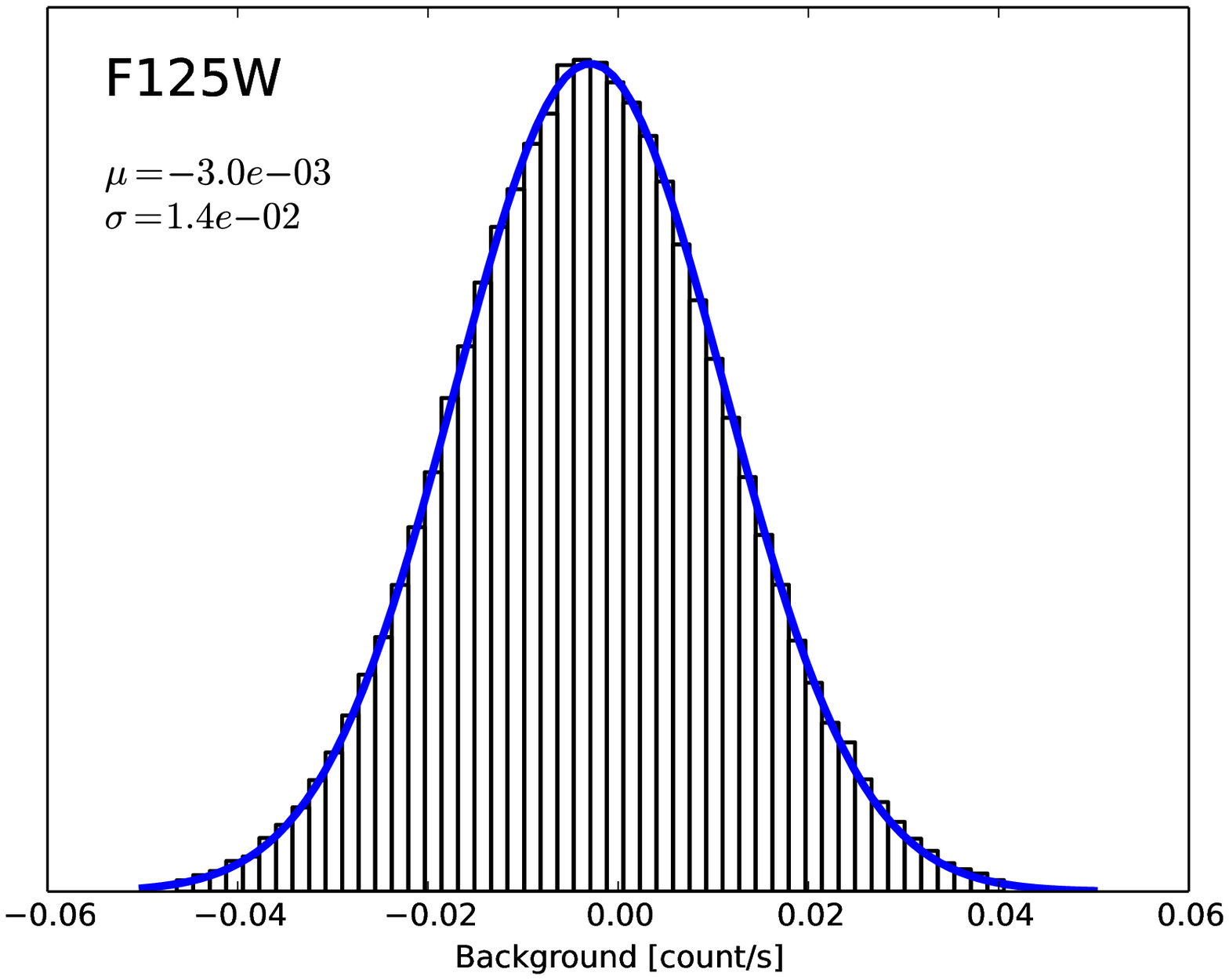}\includegraphics[scale=0.30]{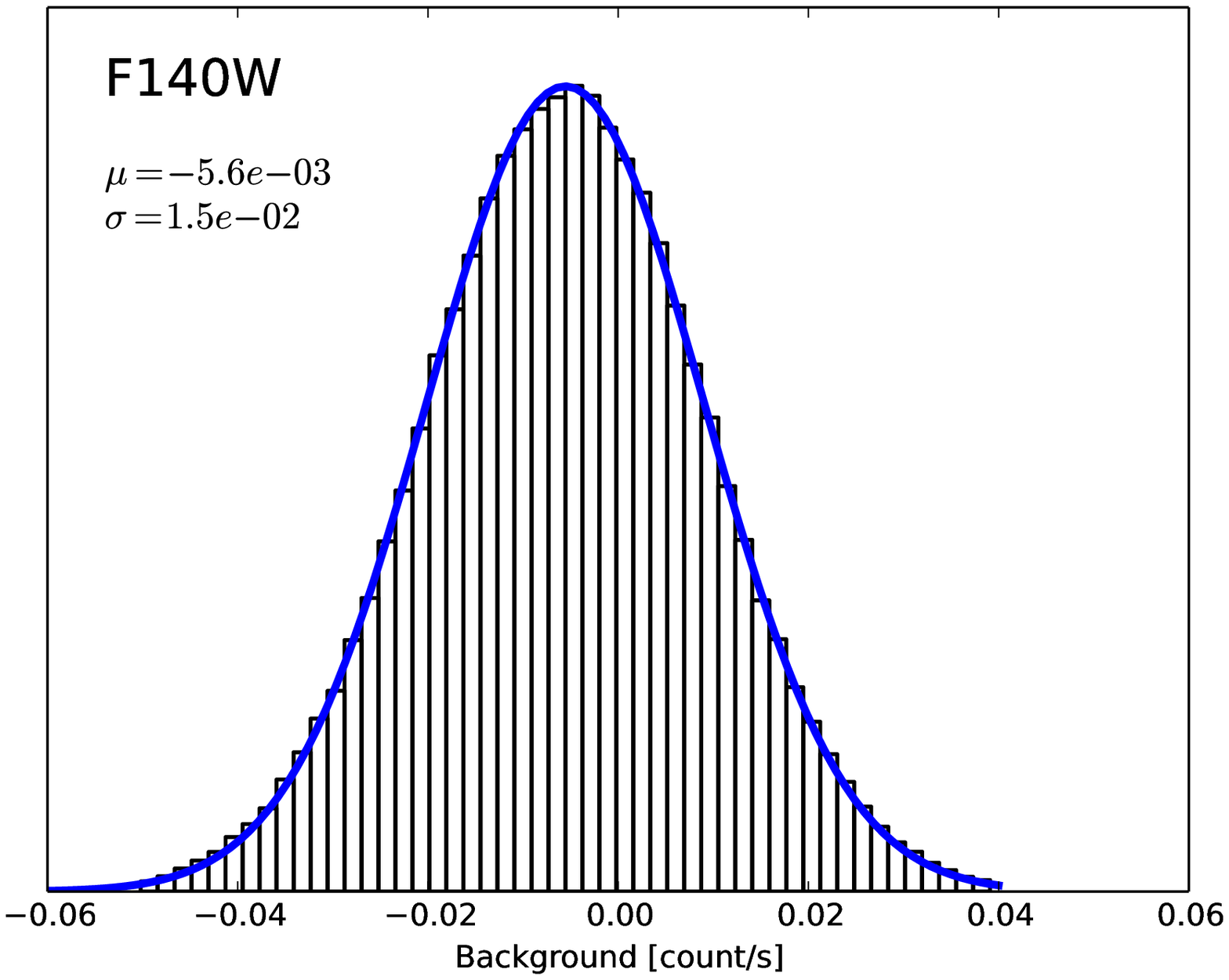}\includegraphics[scale=0.30]{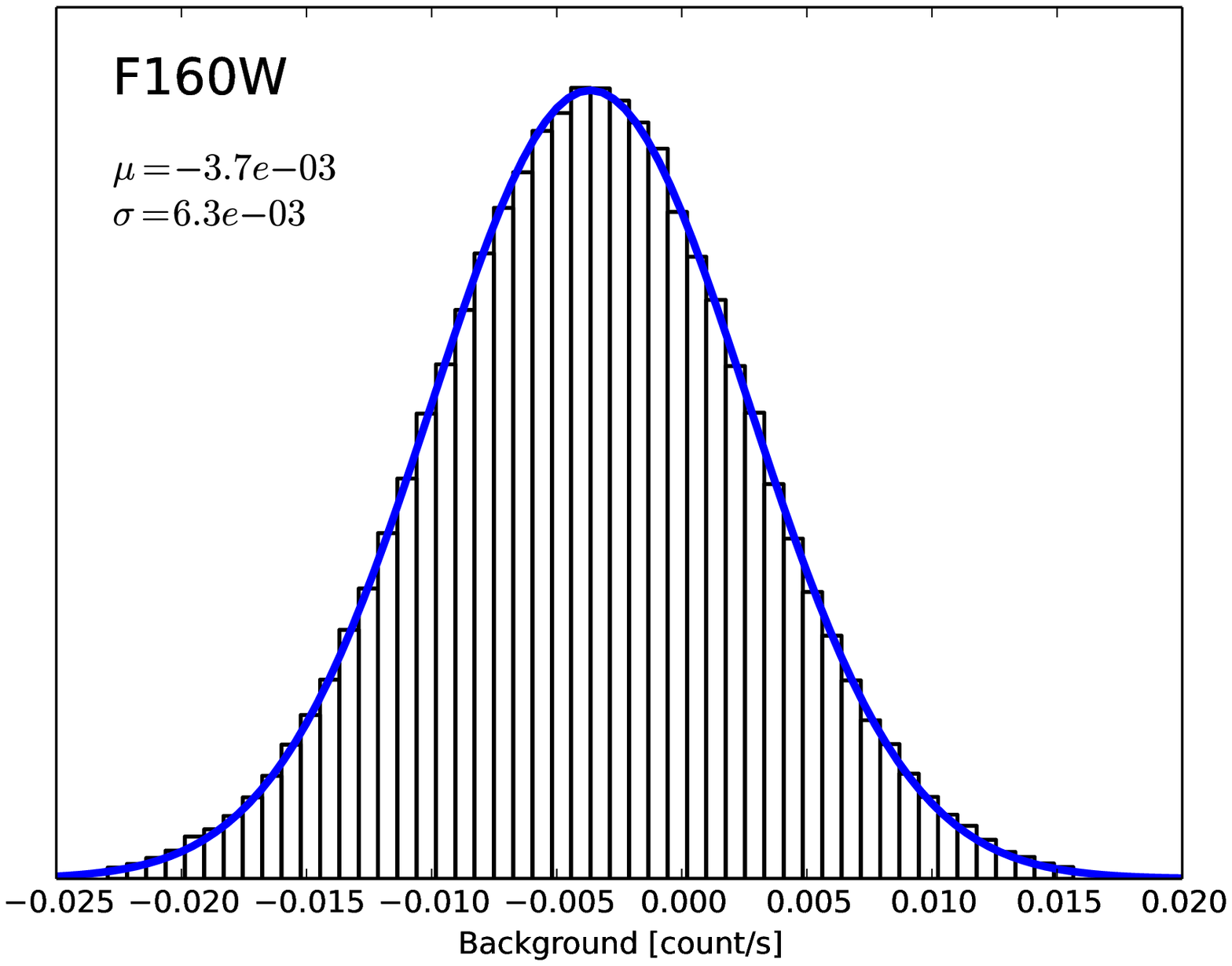}
\caption{Background histograms of the nine filters. Blue lines are the best Gaussian fit to the data, whose mean and standard deviation are shown in the inset.}
\label{fig_histograms}
\end{figure*}

Background histograms are plotted in Fig. \ref{fig_histograms} along with the estimated $\mu$ and $\sigma$. As the light contaminating the borders of the image in the F814W filter was more spread, the blank areas were smaller and the corresponding histogram has less counts. As expected, the obtained values are negative, that is, the background in the original images was oversubtracted by the main pipeline. We also used the blank regions to compute the surface brightness limit for each of the filters, following the technique described by \cite{roman2020}. In short, we estimated the rms in boxes of $10\times 10$ arcsec$^2$, randomly distributed across the blank regions, and measured the surface brightness of the detections at the $3\sigma$ level. The resulting limits, listed in Table \ref{table_data}, range from 28.7 to 30.4 mag arcsec$^{-2}$.\\


\subsection{Photometric Membership via Machine Learning}
\label{phot_memb_ML}

We aimed to select galaxy cluster members based solely on photometric
information of the galaxies lying along the line of sight of our clusters.
For that goal, we employed a machine learning method to estimate a cluster
membership probability, from which we obtained a membership classification.
That was previously shown to be efficient for low-z clusters ($z \le 0.045$)
by \citet{lop20}. Here we extended this approach to higher-z systems ($z < 1$). 

\subsubsection{Spectroscopic Cluster Membership}
\label{spec_members}

To adopt a machine learning approach we first need to know which galaxies
are true members and interlopers along the line of sight of a subset of
clusters. This data set can be used for training and validation purposes. We used a sample of galaxies with spectroscopic redshifts compiled by the CLASH collaboration. These spectra were gathered from different sources: a) the CLASH-VLT program \citep{rosati2014}, a VIMOS large program that conducted an extensive spectroscopic campaign on 13 of the CLASH clusters with 95\% of completeness up to magnitude $m_R=24$ AB \citep{molino2017}; the NASA/IPAC Extragalactic Database\footnote{https://ned.ipac.caltech.edu/}; and previous works as \cite{sand2008,smith2009,stern2010}. \\

The first step before spectroscopically selecting member galaxies is to derive a refined spectroscopic redshift for the CLASH clusters. We do so to have a uniform determination of the cluster's redshifts. We computed their redshift with the gap technique described in \cite{kat96}, but using a density gap \citep{ada98, lop07, lop09} that scales with the number of galaxies available. The gap technique is used to identify groups in redshift space. We applied it to all galaxies within 0.50 h$^{-1}$ Mpc of the cluster centre. The cluster redshift is then given by the biweight estimate \citep{bee90} of the galaxy redshifts of the chosen group. Then we proceed as described below.\\

To select members and exclude interlopers we applied the ``shifting gapper''
procedure \citep{fad96}. For each cluster, we start selecting all galaxies
within 2.50 h$^{-1}$ Mpc and showing a velocity
offset of $| \Delta_v | \le 5000$ km s$^{-1}$. The ``shifting gapper''
procedure is based on the application of the gap-technique in radial bins,
starting in the cluster center. The bin size is 0.42 h$^{-1}$ Mpc or larger to force the selection of at least 15 galaxies.
Those not associated with the main body of the cluster are eliminated. Then, the ``shifting gapper'' is iteratively applied again until the number of cluster members is stable. It is
important to stress that this method makes no hypotheses about the dynamical status
of the cluster. Once we have a member list we obtain estimates of velocity
dispersion ($\sigma_P$), as well as of the physical
radius and mass ($R_{500}$, $R_{200}$, $M_{500}$ and $M_{200}$).\\ 

Our ``shifting gapper'' approach is similar, but not identical to \citet{fad96}.
The most important difference is the adoption of a variable gap, instead
of a fixed one. The variable gap scales with the number of galaxies in the
cluster region and the velocity difference of those belonging to the cluster. Note also that galaxies projected along the line of sight with $| \Delta_v | > 5000$ km s$^{-1}$ are automatically named interlopers (not members). Further details on the spectroscopic membership selection can be found in \citet{lop09,lop14}.\\

After these steps, the sample we had for training and evaluation purposes comprised 1057 galaxies, within the regions of 18 CLASH clusters (spanning a wide redshift range, $0.0792 < z \ < 0.8950$). However, when we performed the final training on our data (Sect. \ref{sect_ml}), we verified that the best results were obtained by selecting only galaxies within 1.50 h$^{-1}$ Mpc of the cluster centre and with 15 $\le$ F814w $\le$ 25. Hence, our final data set comprised 927 galaxies with spectroscopic redshifts.\\


\subsubsection{Machine Learning}\label{sect_ml}

In \citet{lop20} we tested the performance of eighteen machine learning
algorithms, finding six algorithms with superior performance. For the current
data set we found that the {\it Stochastic Gradient Boosting} (GBM) method leads to
slightly better results, which is quantified by two parameters: 
completeness (also known as the "True Positive Rate", TPR or "Sensitivity")
and purity (known as "Precision" or "Positive Predictive Value", PPV). As
described in \citet{geo11}, completeness and purity track the relation between
the sample of objects classified as members and the true population of members.
Completeness is the fraction of true members that are  classified as members,
while purity is the fraction of true members among the objects classified as
members.\\

Gradient boosting is a technique that can be used for regression and classification problems. Analogous to the random forest, the final model is an ensemble of weak prediction models, normally decision trees. However, differently than models based on {\it bagging}, methods in the form of {\it boosting} result in decreased classification bias, instead of variance. The {\it Stochastic Gradient Boosting} is the result of a modification proposed by Friedman (see \citealt{fri02}). He proposed that "at each iteration a subsample of the training data is drawn at random (without replacement) from the full training data set. This randomly selected subsample is then used in place of the full sample to fit the base learner and compute the model update for the current iteration." This randomization process helps improving accuracy, execution speed, as well as increasing robustness against overcapacity of the base learner.\\

Galaxy members and interlopers projected along the line of sight of clusters generally show different distributions for many parameters, such as magnitudes and colors, but also structural and environmental properties. That is shown in Figure 4 of \citet{lop20}. In the ML terminology those parameters are called features. As in \citet{lop20}, we ranked the features by ``importance'', but also tested the performance of our algorithms with different choices of variables. We finally chose the following parameters for training and evaluating our results: $\Delta$(F435W-F814W), $\Delta$(F606W-F814W), $\Delta$(F105W-F140W), $\Delta$(F814W-F125W)
$\Delta$(F814W-F140W), LOG $\Sigma_5$, and $\Delta z_{\text{phot}}$. The
$\Delta$ stands for an offset relative to the mean magnitude, color or
cluster redshift. $\Sigma_5$ is the projected galaxy density to the fifth neighbor, in units of galaxies/Mpc$^2$ (a measure of the local environment). As the clusters in our sample span a large redshift range  ($0.0792 < z \ < 0.8950$),
we do not use apparent magnitudes or observed colors. We also decided to avoid
the use of the absolute magnitudes and rest-frame colors, which are subject
to large uncertainties due to the $k$ and $e-$corrections, as well as on the
photo-z precision.\\

\begin{figure}[]
\centering
\includegraphics[width=9cm]{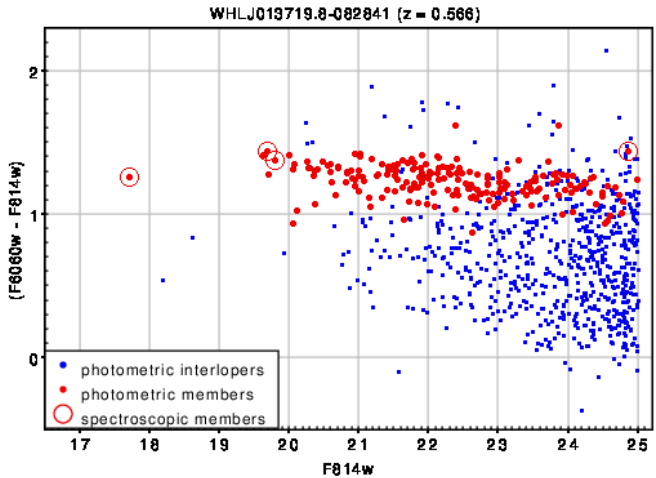}
\caption{Color-magnitude diagram for the WHL0137 cluster with the members (red) identified by the machine learning algorithm, and the interlopers (blue). Members with spectroscopic redshifts are marked with circles.}
\label{fig_CMD}
\end{figure}

After training our method with the parameters above we obtained high values of
completeness (C) and purity (P), C $ = 93.5\% \pm 2.4\%$ and
P $ = 85.7\% \pm 3.1\%$. This method was then applied to all galaxies in the regions of the 25 CLASH clusters, as well as for WHL0137, for the current work. We restricted this work to galaxies with 15 $\le$ F814w $\le$ 25. For the current data set we found the ML classification has the best performance in this magnitude range. We show in Fig. \ref{fig_CMD} the resulting color-magnitude diagram for this cluster, where the red sequence of the cluster is clearly visible. We searched NED\footnote{http://ned.ipac.caltech.edu/} for sources with spectroscopic redshift near WHL0137 and we found just seven objects at this cluster's redshift (maximum velocity offset $<$ 1300 km s$^{-1}$), being four of them within the HST footprint. One of these four galaxies is the BCG, observed by \citet{plk16}, while the remaining three were observed by the Sloan Digital Sky Survey (SDSS\footnote{https://www.sdss.org/dr13/}, \citealt{alb17}). These four galaxies with confirmed spectroscopic membership are indicated in Fig. 3 with large red circles.\\

\subsection{Calculation of the ICL fraction and error}\label{sect_ICLfanderror}

Once the galaxy cluster members are photometrically identified, we readded their CHEF models into the image just containing ICL, to obtain the total cluster light distribution. The profiles of the total cluster light and the ICL, centered on the peak of the ICL, are compared. We defined the maximum radius of the ICL as that where the ICL profile is minimum, taking into account the background level and noise. This minimum always appears, mainly due to technical or observational reasons: the ACS/WFC and WFC3/UV fields of view are not large and the light from the borders of the image contaminate the profiles at large radius, making them artificially increase. This minimum certainly will not represent the outer, physical limit of the ICL, but it is the limit that the own data impose.\\

The error in the final ICL fractions comes from three different sources: the error associated with the cluster membership estimation, the photometric error, and the error associated to CICLE's determination of the BCG limits. The latter is estimated through mock images that mimic the observational characteristics of the BCG+ICL system: the profiles and magnitudes of the BCG and the ICL, and the level of noise of the real images.\\

For further details on CICLE and the photometric membership through ML, as well as tests of their accuracy, we refer the reader to \cite{jimenez-teja2016} and \cite{lop20}.\\ 

\section{Results}\label{sect_results}

\begin{figure*}[]
\centering
\includegraphics[width=9cm]{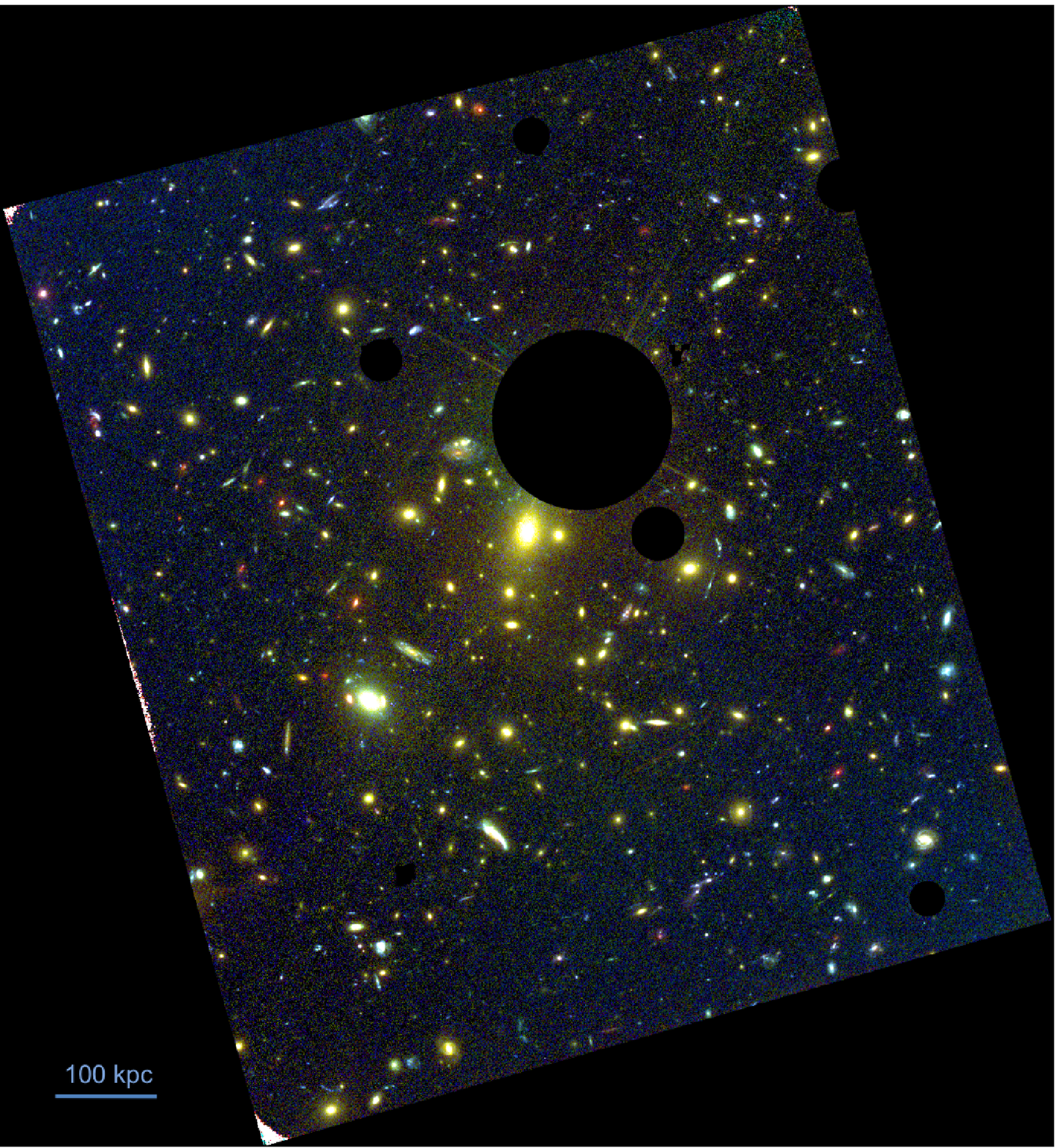}\includegraphics[width=9cm]{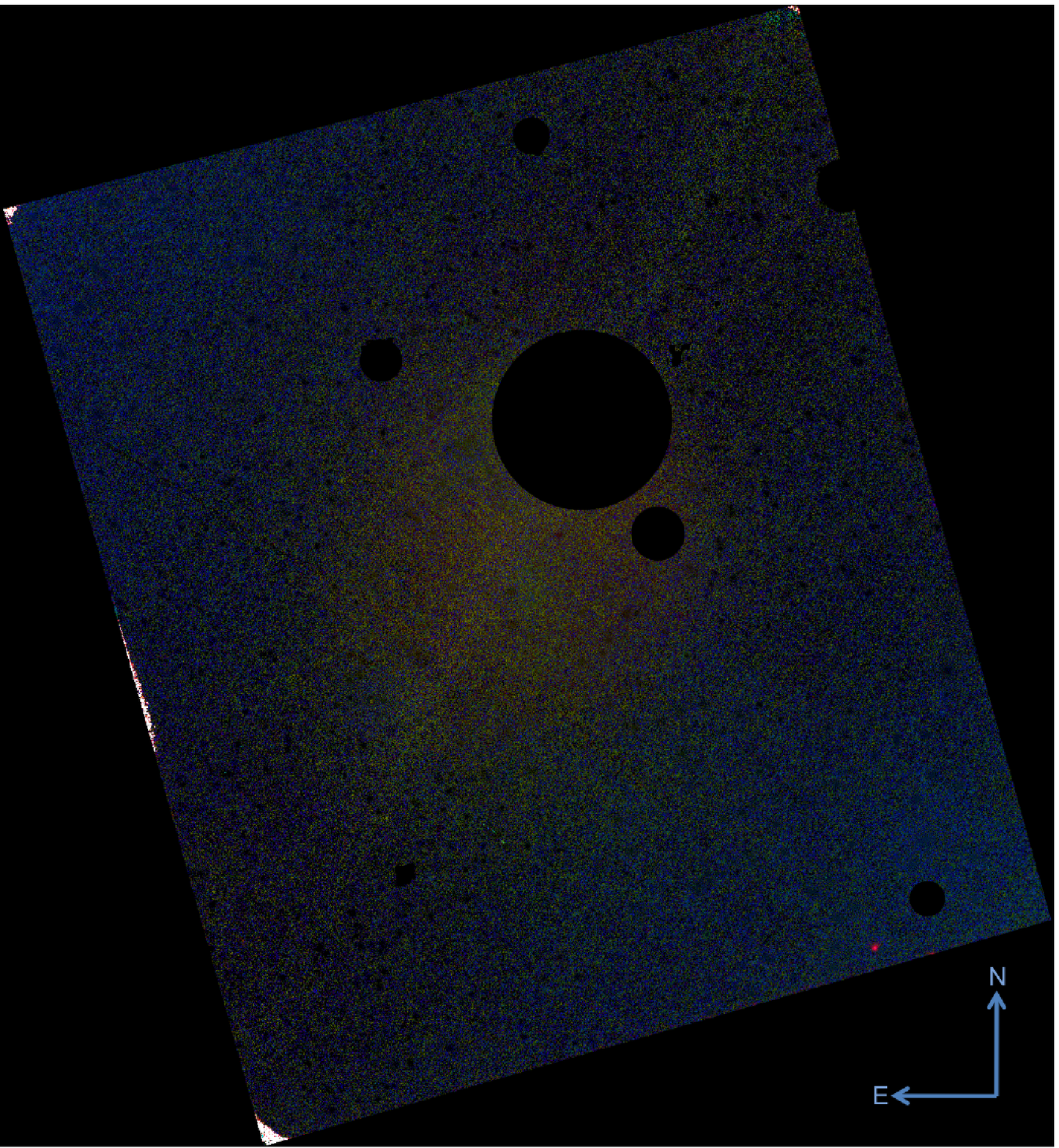}
\caption{Color composition of the original image (left) and the ICL (right) calculated by CICLE.}
\label{fig_original_and_ICL}
\end{figure*}

Here we describe the ICL fractions and intermediate products calculated for WHL0137 in the nine HST bands available: the four optical ACS/WFC filters F435W, F475W, F606W, and F814W, and the five WFC3/IR filters F105W, F110W, F125W, F140W, and F160W. First, as CHEFs can only model galaxies, we masked the stars manually with circular masks, previously to applying CICLE. Extra care was taken to mask the bright star close to the BGG, to avoid any contamination in the ICL maps (Fig. \ref{fig_original_and_ICL}). Please note that these masked pixels were no longer used in any of the following calculations. \\

We then applied CICLE (Sect. \ref{sect_cicle}) to these masked images to obtain images just composed by ICL and background. For each filter, a constant background was estimated to refine the original background subtracted by the HST pipeline during the reduction and coadding processes (see Sect. \ref{sect_bg}). As observed in Fig. \ref{fig_histograms}, the values found for the refined background are all negative, meaning that the original HST pipeline, although optimized for more general science cases, tends to misidentify low surface brightness sources (as the ICL) with background, thus overestimating the latter. These constant backgrounds were subtracted from the star-masked, galaxy-removed images to get the final ICL maps. We created colored images of both the original data and the ICL map (Fig. \ref{fig_original_and_ICL}) using the images in the nine filters: we firstly combined the bands in groups of three using Swarp \citep{bertin2002} and later built the colour image from the three composite images using Stiff \citep{bertin2012}.  The resulting ICL surface is smooth and we do not observe residuals from the galaxy fitting. Only extremely faint structures are present in the ICL map, mainly a crossed-hatched pattern characteristic due to correlated noise in HST images (already noticed by \cite{jee2010} and \cite{jimenez-teja2016}) and very weak residuals from the diffraction cross of the brightest star. Compared to the error in the ICL fractions (see Sect. \ref{sect_ICLfanderror}), this contribution is negligible. We also show a colored composition of the original data, with the same color scale as the ICL, for illustration purposes. \\

We identified 215 galaxies as cluster members in the ACS/WFC field of view of WHL0137 (Sect. \ref{phot_memb_ML}). The corresponding 215 CHEF models were reinserted in the ICL maps, thus keeping the regions corresponding to the cluster members intact, exactly the same as in the original data. Comparing the profiles of the ICL and the total cluster (galaxies plus ICL) at the different wavelengths, we estimated the ICL fractions listed in Table \ref{table_ICLfractions}. We remark here that the small radius found for the ICL in the filter F435W is likely not physical, but instead strongly constrained by the low signal-to-noise of this bluer band.  No k- or evolutionary correction is applied to our photometry and, thus, to our ICL fractions. Since these corrections critically depend on the assumptions made on the spectral energy distributions (SEDs) of the stellar populations, whose nature is a priori unknown for the ICL, several works in the literature usually assume the mean parameters of the BCG as those of the ICL and the total cluster light \citep[e.g., ][]{krick2007,burke2012}. Under this assumption, the k+e correction would be cancelled out in the ICL fractions. \\

\begin{figure*}[]
\centering
\includegraphics[width=18cm]{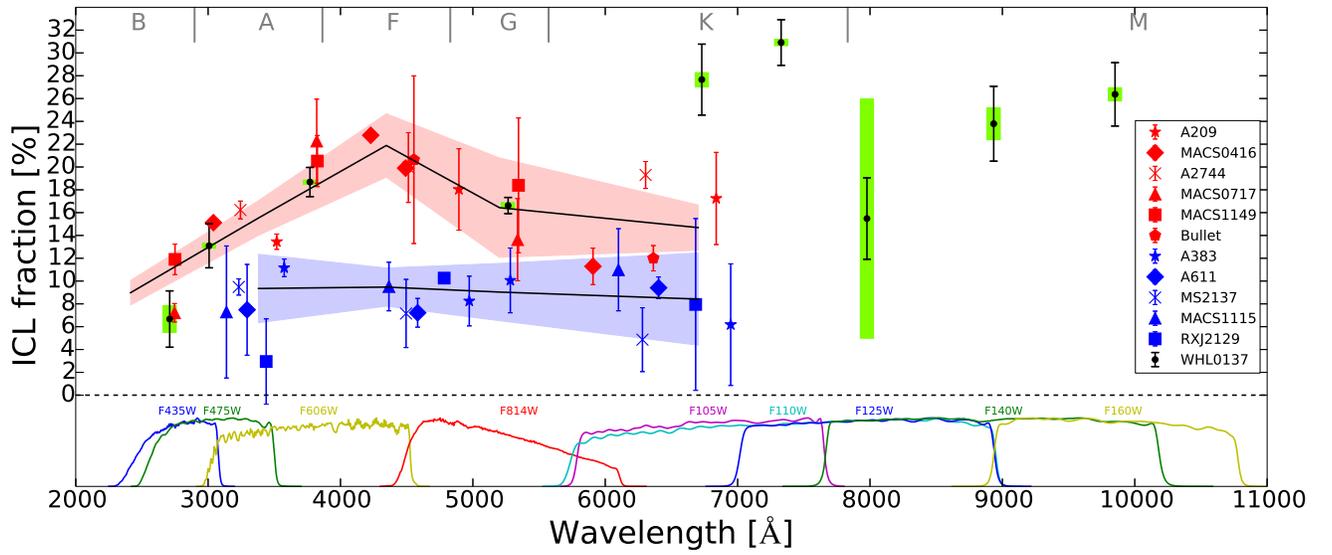}
\caption{ICL fractions of WHL0137 (black points) compared to those found by \cite{jimenez-teja2018} for a sample of merging (red) and relaxed (blue) clusters. All measurements are in the rest-frame. Green rectangles represent the noise weighted errors associated with the ICL fractions of WHL0137, as a way to indicate the impact of the quality of the data in the final ICL fractions. The huge rectangle in the F125W filter makes this measurement uncertain, while the smaller rectangles in filters F475W, F606W, F814W, and F110 mean that these are highly reliable. The red (blue) shadowed region is the error weighted average of the ICL fractions for the merging (relaxed) subsamples. Gray letters on the top split the wavelength range according to the emission peaks of the different stellar types. We show at the bottom the transmission curves of the HST filters analysed in this work, at the redshift of WHL0137 ($z=0.566$), for illustration purposes.}
\label{fig_ICLfractions}
\end{figure*}

Figure \ref{fig_ICLfractions} shows these ICL fractions, overplotted over those found by \cite{jimenez-teja2018} in the optical for a sample of merging (red) and relaxed (blue) clusters, all of them observed by the HST telescope within the frame of the FF and CLASH programs. All ICL fractions are in the rest-frame. As explained in \cite{jimenez-teja2018}, the subsample of relaxed clusters had nearly constant ICL fractions at different optical wavelengths, indicative of an ICL that is just passively fed by the dynamical friction suffered by the galaxy members that evolve towards the center of the cluster potential. As a result, the SEDs of the stellar populations in the ICL and the cluster galaxies have a similar shape (constant ICL fractions). However, the subsample of merging clusters is characterized by: a) an enhancement in the amount of ICL detected (7--23\% for the merging subsample versus 2--11\% for the relaxed systems), and b) an excess in the ICL fraction measured at intermediate, optical wavelengths (from 3800 to 4800\AA, approximately). This was explained invoking a significant amount of ICL stars that was injected more recently, triggered by the merger: the tidal stripping of massive galaxies and the total disruption of dwarf galaxies. Although we do not have previous measurements to compare our result in the IR, we observe that our IR ICL fractions are significantly higher than those in the optical, which will be discussed in Sect. \ref{sect_discussion}.\\

\begin{table}[]
\caption{ICL fraction and error, for the nine HST filters analyzed. The radius where the ICL profile reaches its minimum value is also shown. As described in Sect. \ref{sect_ICLfanderror}, the profile of the ICL typically decreases until a minimum point that is mainly defined by the background level and noise. We measured the ICL fraction up to this point, which may not represent the real physical limit of the ICL, but the limit of the data. }\label{table_ICLfractions}
\centering
\begin{tabular}{ccc}
   Filter & ICL fraction & $r$ \\
   & [\%] & [kpc]\\
\hline
F435W & $6.67\pm2.47$ & 44.2\\
F475W & $13.09\pm1.93$ & 152.3\\
F606W & $18.68\pm1.28$ & 169.7\\
F814W & $16.62\pm0.71$ & 403.7\\
F105W & $27.66\pm0.50$ & 408.2\\
F110W & $30.91\pm0.31$ & 407.8\\
F125W & $15.48\pm3.57$\footnote{This value should be taken with caution, given the high level of noise found for the F125W filter.} & 372.1\\
F140W & $23.79\pm0.59$ & 360.1\\
F160W & $26.37\pm0.54$ & 407.7\\
\hline
\end{tabular}
\end{table}

\section{Possible systematics} \label{sect_systematics}

Previous to the interpretation of the ICL fractions found, we need to discuss and rule out (if this is the case) the different sources of bias for our data and measurements. A very nice description of several systematics that can have an impact on low surface brightness studies using HST data can be found in \citet{borlaff2019}. Here, we will focus on five of them: persistence, flat-fielding, zodiacal light and earthshine, atmospheric He I $\lambda$1.083 $\mu$m emission, and dust.\\

\subsection{Persistence}

\begin{figure*}[]
\centering
\includegraphics[width=18cm]{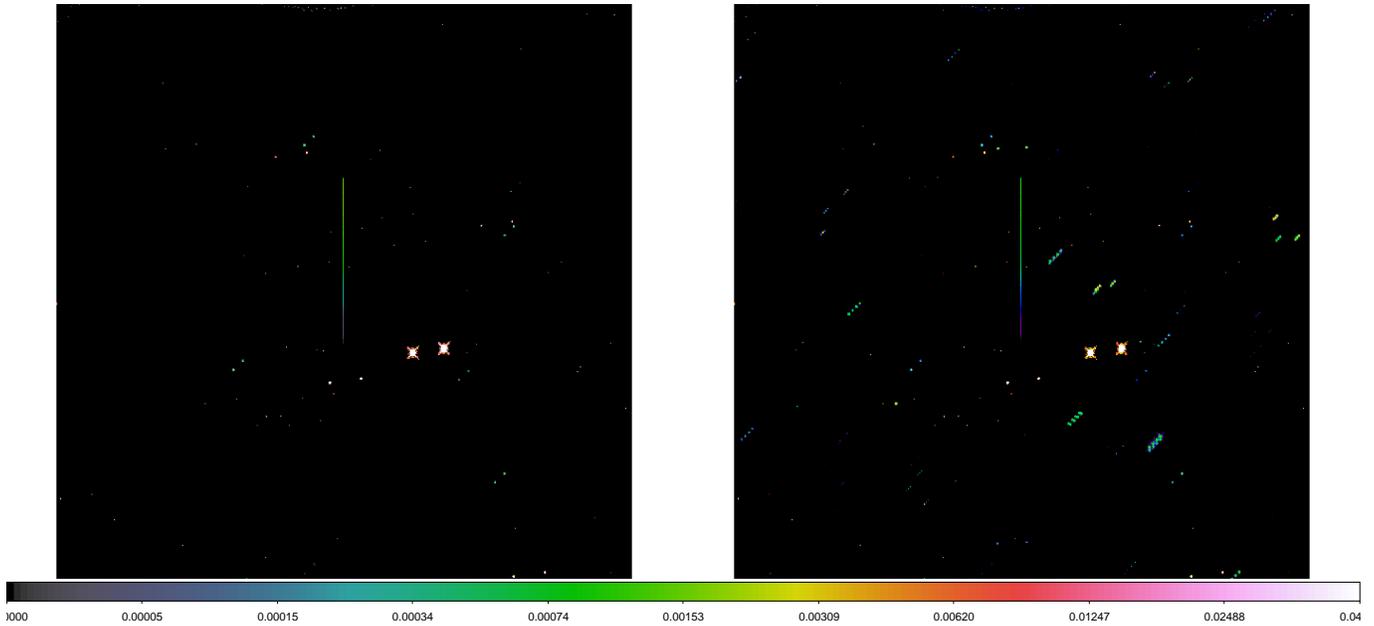}
\caption{Persistence models calculated with a lookback time of 16 hours (left) and 96 hours (right), using the code developed by \cite{long2015}}.
\label{fig_persistence}
\end{figure*}

It is well-known that HgCdTe IR array detectors, as the WFC3, can suffer from persistence. When pixels in a detector are exposed to very bright sources of light (namely, more than about half of the full potential well of the pixel), after images can appear in the following observations during a period of time \citep{long2012}. This afterglow can certainly have an impact on any measurement made over a low surface brightness object and, therefore, it must be corrected or quantified. The WFC3/IR observations of WHL0137 already had persistence masks applied by the RELICS collaboration (see Sect. \ref{sect_data}). However, these masks were calculated using a default lookback time of 16 hours, which, according to \cite{borlaff2019}, is insufficient. A minimum time of 96 hours is required to check all the previous observations to those of WHL0137 and compute the persistence masks. Thus, we used the publicly available code developed by \cite{long2015} to identify the pixels affected by persistence\footnote{https://github.com/kslong/Persistence/wiki} and create 96-h persistence models. In Fig. \ref{fig_persistence}, we show as an example, the 16-h and 96-h persistence models for the F160W filter. We observe that pixels with the brightest afterglow were already included in the 16-h model. However, a minor contribution coming from a few, fainter pixels was missed in the 16-h model. \\

We checked that these missed pixels had values that ranged from 0.01 to 0.03 e$^-$/s for the five IR filters considered. Specifically, the number (and percentage) of missed pixels is reported in Table \ref{table_persistence}. We then calculated the extra amount of flux added by these pixels and compared it with the measured ICL fluxes and ICL fractions, to quantify the impact on the results.\\

\begin{table*}
\caption{Impact of the pixels affected by persistence in the 96-h models and missed by the 16-h models on the ICL measurements: number of missed pixels and percentage with respect the total number of pixels in the image, percentage of flux artificially added to the ICL by these missed pixels, and error induced in the ICL fraction.}\label{table_persistence}
\centering
\begin{tabular}{cccc}
   Filter & \# missed pixels (percentage) & Flux added to the ICL & ICL fraction error \\
   & [pixels (\%)]& [\%] & [\%] \\
\hline
F105W & 278 (0.027 \%) & 4.0e-4 & 8.3e-5\\
F110W & 493 (0.048 \%) & 2.1e-4 & 4.9e-5\\
F125W & 2262 (0.220 \%) & 1.3e-2 & 1.7e-3\\
F140W & 1034 (0.100 \%) & 9.0e-4 & 1.6e-4\\
F160W & 6120 (0.595 \%) & 1.0e-2 & 2.0e-3\\
\hline
\end{tabular}
\end{table*}

According to Table \ref{table_persistence}, we observe that the most affected bands are F125W and F160W, with a maximum of $\sim$ 0.6 \% affected by persistence and not included in the 16-h models. Assuming the worst case scenario (that all missed pixels with persistence are located in the region where the ICL is detected), these missed pixels increased the ICL flux, at most, a 1.3e-2 \% (F125W band). We calculated new ICL fractions subtracting the extra flux added by the missed pixels. When we compared them with the original ICL fractions, we found a maximum difference of 0.002 \%, which is safely folded by the ICL fraction errors reported in Table \ref{table_ICLfractions} where the minimum error found is 0.31\% for the F110W filter. Therefore, we conclude that, although 96-h persistence models can certainly have an impact on ICL measurements when smaller areas are considered, their effect is dramatically reduced when calculations are made over large areas, as it is our case. Moreover, as we measure ICL fractions and persistence affects to both the ICL and the total cluster flux, its effect is somehow attenuated by the fraction. \\

\subsection{Flat-fielding}

\begin{figure*}[]
\centering
\includegraphics[width=8.5cm]{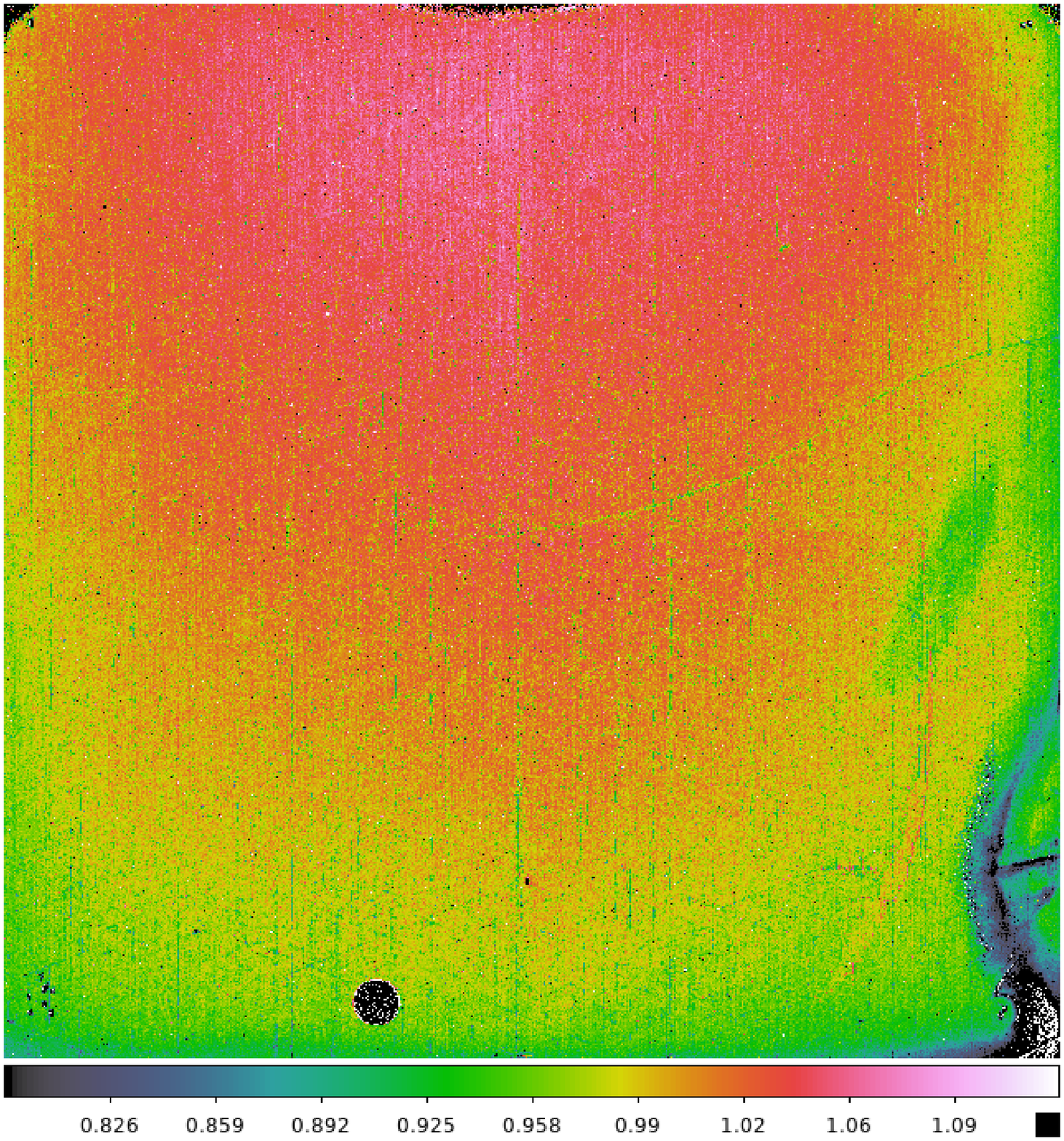}\hspace{1cm}\includegraphics[width=8.5cm]{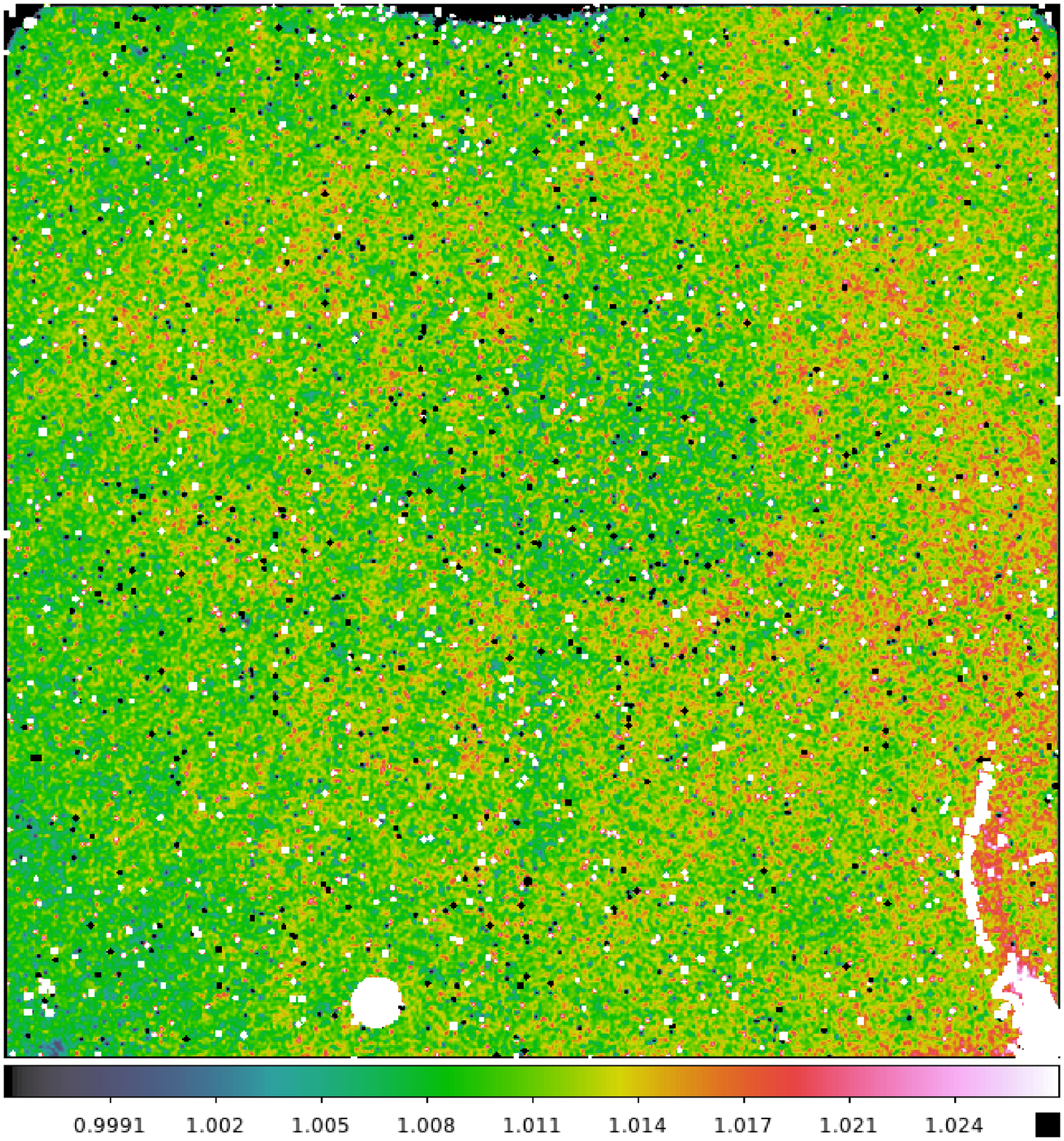}
\caption{Left: sky-flat field computed for the F160W band of WHL0137, following the algorithms described by \cite{pirzkal2011} and \cite{borlaff2019}. Right: Ratio of our sky-flat field to the official MAST flat, with the contrast enhanced using a 3 pixel-wide Gaussian}.
\label{fig_flats}
\end{figure*}

As described in Sect. \ref{sect_data}, flat-field images of WHL0137 were produced using the HST standard calibration pipelines CALACS and CALWF3. Nevertheless, as shown by \citet{borlaff2019}, these pipelines might not be optimal for low surface brightness objects or features, which are often undetected or misidentified with noise. The alternative algorithms proposed by \citet{borlaff2019} and \citet{pirzkal2011} found a maximum difference between the official MAST and the new flat-field images of the Hubble Ultra Deep Field of 1\%. We proceed in a similar way to calculate new, sky-flat fields for WHL0137. For each one of our nine optical and IR filters, we downloaded the ``flt'' (for WFC3) and ``flc'' (for ACS) files of all the observations made ``close'' in time with those of WHL0137. The time window considered depended on the number of observations made with each filter with an exposure time higher than 300 seconds, as described by \citet{pirzkal2011}. The idea is having a minimum of 200 images to build each flat-field image, to have enough statistics. We visually inspected and removed those observations that could compromise the flat due to the presence of extended, low surface brightness emissions or poor quality of the data. Then, we ran NoiseChisel \citep{akhlaghi2015} to efficiently mask all the objects in every flt image and we multiplied them back by their corresponding official MAST flat fields. Finally, we normalized and combined these masked images as described in \citet{pirzkal2011}. For the sake of illustration, we show in Fig. \ref{fig_flats} the resulting sky-flat field in the filter F160W and its ratio with the official MAST flat (smoothed with a 3 pixel-wide Gaussian to enhance the possible differences, as in \citet{borlaff2019}, see their Fig. 4).\\

Later, we applied the new, sky-flat fields to the individual flt images of WHL0137 and we ran Astrodrizzle \citep{koekemoer2002} to align, drizzle, and combine the images, with the same parameters as in the original pipeline, which are described in the corresponding ``mdz'' table. We use these new reduced images to run CICLE and calculate new ICL fractions, which are reported in Table \ref{table_flats} along with the difference with the original ICL fractions listed in Table \ref{table_ICLfractions}. As observed, the new flats do not induce a significant bias in the resulting ICL fractions. Moreover, the difference is smaller than the errors reported in Table \ref{table_ICLfractions}, which means that the possible bias in the official MAST flat-field images is folded into our error measurement, mainly because our sky background estimation accounts for it. Besides, as in the case of the persistence, the multiplicative effect of the flat is diluted by the use of ICL fractions.\\

\begin{table}
\caption{ICL fractions computed from the images corrected with the sky-flat fields  derived with the algorithms described by \cite{pirzkal2011} and \cite{borlaff2019}. We also show the difference with the ICL fractions computed from the images corrected with the official MAST flats.}\label{table_flats}
\centering
\begin{tabular}{cccc}
   Filter & Sky-flat ICL fractions & Difference \\
   & [\%]& [\%] \\
\hline
F435W & 6.36 & 0.31\\
F475W & 11.83 & 1.26\\
F606W & 18.42 & 0.26\\
F814W & 17.02 & 0.40\\
F105W & 27.72 & 0.06\\
F110W & 30.83 & 0.08\\
F125W & 17.35 & 1.87\\
F140W & 24.15 & 0.36\\
F160W & 26.03 & 0.33\\
\hline
\end{tabular}
\end{table}

\subsection{Zodiacal light and earthshine}

\begin{figure}[]
\centering
\includegraphics[width=9cm]{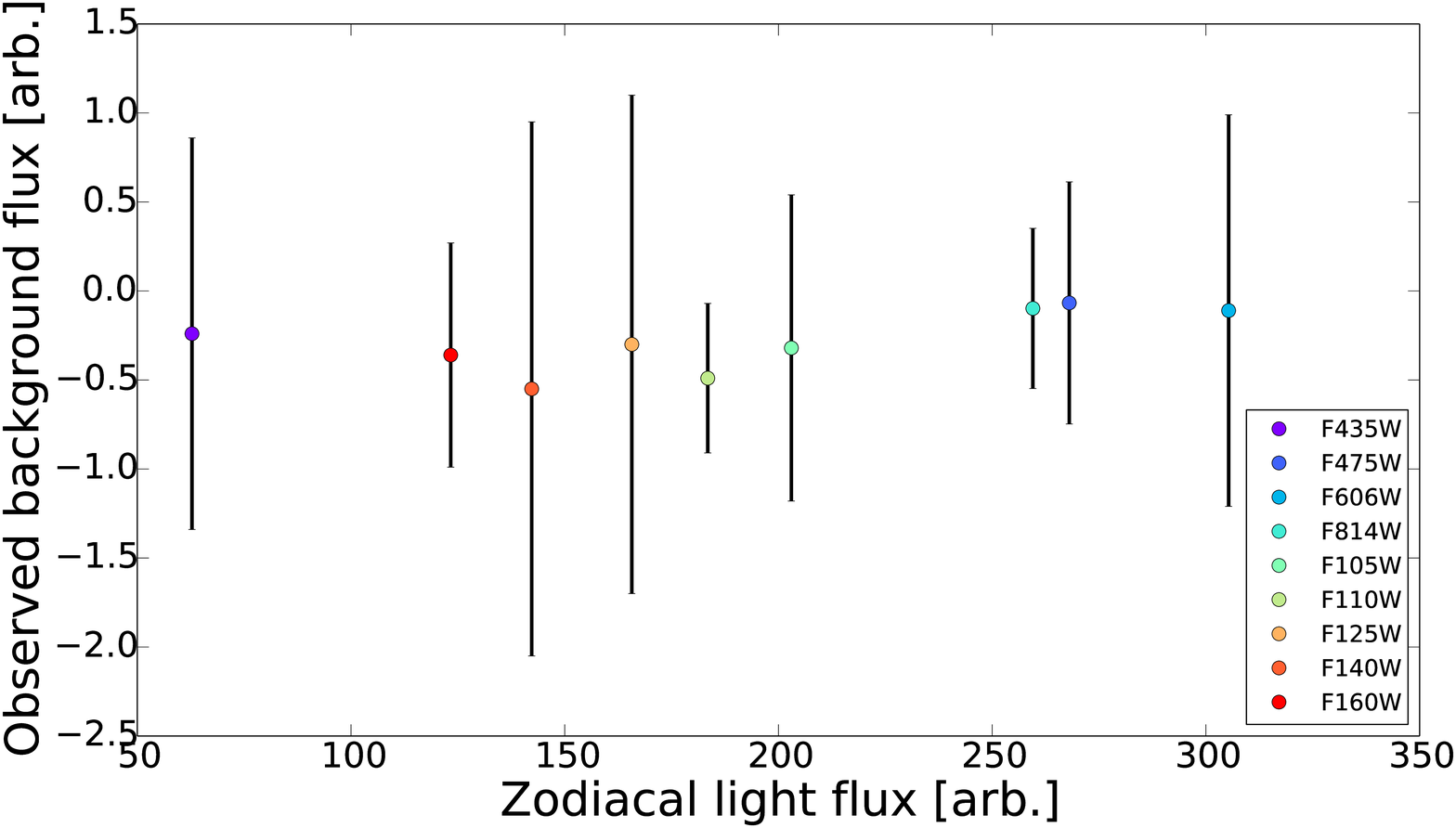}
\caption{Comparison between the refined background and the zodiacal light, for the nine HST bands. Zodiacal light is modelled assuming a solar spectrum.}
\label{fig_ZL}
\end{figure}

 We raise here the question of whether the higher IR values found (in comparison to the optical ones) are the consequence of any systematic or IR contamination.
 In space, the dominant sources of background radiation at shorter IR wavelengths are zodiacal light and earthshine, being F160W the least affected filter\footnote{https://www.stsci.edu/itt/APT\_help/WFC3\_Cycle23/c07\_ir10.html\#364150}. For relatively low ecliptic latitudes as that of WHL0137 (-17.28741720$^{\circ}$), observational strategies as taking the angle between the target and the Sun as viewed by the HST (that is, something equivalent to the solar elongation angle used in planetary sciences) greater than 50$^{\circ}$ or 60$^{\circ}$ diminish the impact of this glowing light in the background. For our WHL0137 WFC3/IR observations, this angle was set to $\sim$59.6$^{\circ}$ for F105W, F140W, and, F160W, and to safer 152.2$^{\circ}$ and 95.9$^{\circ}$ for F110W and F125W, respectively. Additionally, these effects are globally removed in the post-processing  by the standard HST reduction pipelines, but we compared our refined backgrounds with the estimated zodiacal light to check if some contamination could have slipped in. Following a similar approach that \cite{morishita2017}, in Fig. \ref{fig_ZL} we compare our background measurements with the zodiacal light flux density. This latter was calculated at the position of WHL0137 in the dates of the observations, assuming that zodiacal light has a solar spectrum. We do not see any correlation between the refined background and the zodiacal light estimations, which indicates that the effect of the zodiacal light was efficiently removed by the HST reduction pipeline and, thus, it does not impact our ICL fractions.\\

\subsection{Atmospheric He I $\lambda$1.083 $\mu$m emission}

Atmospheric He I $\lambda$1.083 $\mu$m emission can also affect the F105W and F110W data, when the images are taken from low Earth limb angles outside its shadow. However, the FF pipeline (which was also adopted for the RELICS sample) removes the global contribution of this effect, only leaving (if any) small-scale fluctuations in the background \citep{morishita2017}. We assume here that the uncertainty introduced by this faint emission is negligible in comparison with the photometric, geometrical, and cluster membership errors included in the final ICL fractions, so it is not responsible for the higher values found for these two IR filters. This is specially irrelevant in the case of F110W, for which the signal-to-noise is higher given its 2-orbit depth (see Table \ref{table_data}). Additionally, F105W is pretty independent of both F140W and F160W (which are not polluted by the atmospheric He I), and still they show similar results.\\

\subsection{Dust}

Finally, we wondered if the higher IR ICL fractions can be the result of a possible contamination by galactic and intergalactic dust. However, dust reddening is ruled out by two reasons: a) WHL0137 is far from the galactic plane (Galactic coordinates: $l$=155.2885$^{\circ}$, $b$=-68.3898$^{\circ}$ [J2000.0]), so the extinction is low \citep[$E(B-V)=0.0286$, ][]{coe2019}; and b) even if significant, extinction would enter as a multiplicative term in our ICL fractions, so it would be cancelled out. One could ask if intracluster dust may also play an important role in reddening the ICL stars \citep{stickel1998, stickel2002,popescu2000,alberts2021}; in fact, reddened young stars should peak around $1\mu m$. However, it is not expected that dust would survive in the ICM for much more than $10^4-10^5$ yr \citep{draine1979}. Observations of intracluster dust (ICD) so far have been mostly limited to providing upper limits based on extinction of background galaxies \citep[e.g.,][]{gutierrez2014,gutierrez2017}. Even if ICD is present \citep[e.g.][]{longobardi2020}, its contribution is negligible especially at the very near IR used here. Theoretical efforts to justify these upper limits needed to use values for the dust lifetime an order of magnitude higher then what is expected \citep[e.g., ][]{vogelsberger2019,mcKinnon2017}. Even at higher redshifts, ICD should be a negligible component, which is not required to explain the observations of WISE, Spitzer and Herschell satellites \citep[e.g., ][]{alberts2021}. On the other hand, the emission from very numerous fiducial stripped old red stars (K-M) to those wavelengths, where an IR excess is found could be substantial when integrated over the lifetime of the cluster and we discuss this below (Sect. \ref{sect_IRICLfractions}).\\

Another interesting effect that should be considered is the fact that cold gas filaments originated from warm galactic outflows in clusters of galaxies can contain dust that can weather the sputtering if the grain size if small enough \citep{qiu2020}. However, this dust would be confined to filaments spanning scales of tens of kiloparsecs, so its effect is expected to be negligible for a cluster at $z=0.566$.\\

\section{Discussion} \label{sect_discussion}

Now that we have ruled out a possible bias in the results, we analyze the implications of the optical ICL fractions in the dynamical state of the cluster, the higher IR ICL fractions with respect to the optical, and the potential fossil group nature of WHL0137.\\

\subsection{WHL0137 dynamical state}

Cluster WHL0137 is at $z=0.566$, right at the high end of the redshift interval considered in \cite{jimenez-teja2018} ($0.18<z<0.55$). We thus try to correlate the ICL fractions measured for WHL0137 with the previous findings. Comparing the optical ICL fractions  with the shaded red region (which is the error-weighted average of the ICL fractions of the merging clusters), we see in Fig. \ref{fig_ICLfractions} that they are consistent: WHL0137 ICL fractions range from 6\% to 19\% in the optical and we clearly see an excess between 3800--4800 \AA, mainly evidenced by the F606W filter and, partially, by the redder F814W. Thus, our optical ICL fractions alone would indicate that WHL0137 is in an active, merging state. \\

Previous works in the literature were mainly focused on the analysis of X-ray, XMM-Newton data of this cluster. \cite{bartalucci2019} analysed a sample of 75 massive clusters to infer their dynamical state from different  X-ray morphological parameters as the centroid shift (separation between the x-ray peak and the centroid of the x-ray distribution), the concentration, or a combination of both. In the case of WHL0137, these three parameters agree and confirm that it is a disturbed, unrelaxed cluster. Thus, the ICL fractions, based on optical data alone, confirm the previous and independent findings based on X-rays.\\

\subsection{ICL fractions in the IR} \label{sect_IRICLfractions}

The analysis described in \cite{jimenez-teja2018} was based solely on optical imaging, so that we do not have previous analysed data in the IR range to compare to WHL0137. Additionally, the WFC3/IR filters strongly overlap, which complicates the interpretation of the ICL fractions measured in the redder regions of the spectrum. Fig. \ref{fig_ICLfractions} shows higher ICL fractions in the IR regime, compared to the optical wavelengths, except for the F125W filter. As a way to quantify the influence of the different observational characteristics of the images into the final ICL fractions, we estimated the impact of the noise in the BCG+ICL composite surface in the nine HST bands, by denoising the images with a wavelet filter and comparing them with the noisy data. We found that for F125W this level of noise was not adequate ($N<1$), and in fact it was $\sim 25$ times higher (on average) than that of the other IR bands. Both the other IR and visible images had reasonable values for $N$, meaning that the original BCG+ICL surface is comparable with its denoised version: 1.05 (F435W), 4.42 (F475W), 3.46 (F606W), 1.33 (F814W), 2.53 (F105W), 3.55 (F110W), 0.17 (F125W), 1.17 (F140W), 2.4 (F160W). Clearly, the errorbars (which result from the combination of the error in the cluster membership, in the determination of the limits of the BCG, and in the photometric measurements) do not fully include the impact of the noise of the images in the final ICL fractions. In order to quantify and introduce the effect of this parameter into the ICL fraction, we have weighted the original errors with $N$. We represented the noise weighted errors as green rectangles in Fig. \ref{fig_ICLfractions}. These new errors provide valuable information that can be associated to each photometric measurement, since they can be understood as quality flags. Whereas for filters F475W, F606W, F814W, and F110W the green rectangles are very small, reinforcing the reliability of these measurements, the huge rectangle associated to the F125W ICL fraction is compatible with the range of values found for all the ICL fractions, both in the visible and the IR, for both merging and relaxed clusters. As a consequence, the uncertainty is high, so this filter is not retrieving any significant information. Additionally, we note that the width of the F110W makes it include the flux corresponding to the wavelengths covered by both the F105W and F125W, and it still yields an ICL fraction in better agreement with the other IR measurements. For these reasons, we will exclude the F125W ICL fraction in the subsequent discussion.\\ 

Once we ruled out a possible contamination or systematic in the measurements (see Sect. \ref{sect_systematics}), we try to unveil the physical cause of these enhanced IR ICL fractions. Higher ICL fractions in the IR mean a higher presence of the redder (older and/or higher metallicity) stars in the ICL, when compared with the average stellar populations locked up in the cluster galaxies. To help with the interpretation of the results, we added at the top of Fig. \ref{fig_ICLfractions} the wavelength intervals that mainly comprise the emission peaks of the different stellar types in the Main Sequence (B, A, F, G, K, and M). Obviously, even though each band receives the contribution from all stellar types, certain types would have a higher influence. Thus, K- and M-type stars mainly emit in the broad-band filters F105/110/140/160W, at WHL0137 redshift of $z=0.566$. Since the comparison with other works based on photometric data is not straightforward due to the disparity of techniques used (see Sect. \ref{sect_intro}), we turn to spectroscopic studies of the ICL to link the information about the stellar populations of the ICL and their impact on the IR photometry.  \cite{melnick2012} calculated that the ICL of the active cluster RXJ0054.0-2823, at $z\sim 0.29$,  is heavily dominated in mass ($91\pm 3$\%) by old stars ($\sim$10 Gyr), of which a 38$\pm$5\% are metal rich. Moreover, they found that $\sim 8\pm 3$\% of the ICL is composed by intermediate-age ($<6$ Gyr) stars of solar metallicity, along with traces ($\sim$1\%) of very young stars ($<10^7$ yr). As this cluster is $\sim$ 2 Gyr older than WHL0137, assuming a similar merging history, this composition would be reflected in higher ICL fractions for the filters that have a larger impact from the G-, K-, and M-type stars. \cite{adami2016} investigated the stellar populations in the cluster XLSSC 116 ($z\sim 0.53$), which is apparently in an intermediate dynamical stage, with a minor merging ongoing. Using both broad-band images from the CFHT ($u^*$, $g'$, $r'$, $i'$, $z'$, and $K_s$) and integral field spectroscopy from MUSE, they inferred that the ICL spectrum follows an early-type spectrum with an [OII] emission line and a mean age ranging between $\sim 1$ and 6 Gyr. Additionally, the average spectrum of the galaxy members (excluding the BCG) is consistent with that of an elliptical galaxy with prominent absorption lines (e.g. H\&K), a strong Balmer break, but without any significant emission lines. Contrarily, the BCG displays a Sd spectrum with strong emission lines. Although this cluster is very close in redshift to WHL0137, the different dynamical states make a fair comparison difficult. However, a rough conversion of these ICL and galactic spectra into ICL fractions would imply that the ICL would show a larger contribution in IR filters (as compared to the visible). Moreover, the IR ICL fractions should indeed be high taking into account that the BCG and the ICL have similar masses ($7.9\times 10^{10}\, M_{\sun}$ and $5.0\times 10^{10}\, M_{\sun}$, respectively). Finally, \cite{edwards2016} analysed SparsePak integral field spectroscopy of three low-redshift clusters ($z<0.06$) with well-defined dynamical states: Abell 85 (relaxed, cool core cluster), Abell 2475 (intermediate stage, minor merger ongoing), and IIZw108 (unrelaxed, pre-merging or ongoing a major merger). In all three clusters, the ICL hosts much younger and lower-metallicity stars than any region of the BCG. Specifically, in the three clusters the ICL is best fit with a large population of alpha-enhanced old, metal-rich stars like the BCG, but also an additional component of young, metal-poor stars. However, the detailed composition of the ICL differs from cluster to cluster: 
\begin{itemize}
    \item Abell 85: 34\% old, metal-poor stars (13 Gyr, Z=0.006), 35\% young, metal-rich stars (5 Gyr, Z=0.032), and 31\% young, metal-poor stars (5 Gyr, Z=0.006),
    \item Abell 2457: 12\% old, metal-poor stars (12 Gyr, Z=0.006), 37\% old, metal-rich stars (12 Gyr, Z=0.032), and 47\% young, metal-poor stars (5 Gyr, Z=0.006),  
    \item IIZw108: 48\% old, metal-rich stars (12 Gyr, Z=0.032), and 55\% young, metal-poor stars (3 Gyr, Z=0.006) 
\end{itemize}
\noindent We can see how the ratio of young and metal-poor population increases as the cluster is more perturbed, which could be seen as enhanced ICL fractions in the bluer filters as compared to more relaxed systems. However, an old and/or higher-metallicity population is ubiquitous independently of the dynamical stage, which would be mostly seen in the IR filters. Although these three clusters are $\sim 4.7$ Gyr older than WHL0137, these redder populations still would leave a higher imprint in the IR wavelengths.\\

We thus speculate that these higher IR ICL fractions, joint with the enhanced ICL fractions in the visible (as compared with the relaxed clusters) can indeed be the reflection of mixed stellar components with two distinct origins: a) a significant amount of bluer stars, probably consequence of the active stage of WHL0137,  whose main mechanisms of injection would be the tidal stripping of infalling galaxies, high speed encounters, and the tidal disruption of dwarf galaxies; and b) a redder population, possibly linked to the assembling of the BCG at $z>1$, the dynamical friction of the extant passive cluster population (ongoing stripping), the aging of stars placed into the ICL by past mergers, and the accretion of ICL pre-processed in infalling groups. In fact, it has been observed that preprocessing in infalling groups might play an important role in the accretion of ICL during the merger. Gravitational interactions within group members can strip material (stars and gas) into the intragroup space, which would be a source of redder stars for the ICL once the group is accreted by the main cluster, as observed, for instance, in Abell 3888 \citep{krick2006} and Virgo \citep{mihos2017}. Simultaneously, these tidal interactions can weaken the potential wells of the group members, favouring ram-pressure stripping by the intracluster medium which can produce a diffuse H$\alpha$ emission and even an associated HI component \citep[see, for instance, the case of Abell 1367, ][]{gavazzi2003,cortese2006}. In this cluster was also observed an inusual density of metal-rich, star-forming dwarf galaxies, in the region of the infalling group, which could end up disrupted into the ICL eventually. These dwarfs and the diffuse H$\alpha$ emission would also contribute to the redder color of the ICL. Nevertheless, a detailed study of the ICL SEDs and colors needs to be completed before any formal conclusion can be reached on the different scenarios, which will be the subject of a forthcoming paper.\\

Another interesting possibility would be the presence of emission-line nebulae, especially in the case of cooling flow clusters \citep{mcdonald2010}. For instance, a H$\alpha$-emitting giant nebula of 80 kpc × 55 kpc is observed around NCG 1275, the BCG in Perseus cluster, consisting on a complex thread-like filamentary structure \citep{gendron-marsolais2018}. Given the higher redshift of WHL0137, such a resolved structure would not be visible, but it may imprint its signature on the ICL fractions. 


\subsection{Is WHL0137 a fossil group progenitor?}

Fossil groups (FGs) are defined as galaxy associations characterized by single, elliptical, dominant galaxy whose next brightest companion within a certain radius (0.5$R_{200}$, that is, half the virial radius) has a difference in magnitude higher than 2.0 mag in the $r$-band \citep{jones2003}. An alternative definition was provided by \cite{dariush2010}, using the magnitude gap between the BCG and the fourth-ranked galaxy instead, and requiring it to be higher than 2.5 mag. Additionally, FGs are characterized by an extended and luminous halo emission in X-rays, typically $L_{X,bol}\geq 10^{42}h_{50}^{-2}$ erg s$^{-1}$. To reach such a large magnitude gap between the BCG and the second-ranked galaxy ($\Delta m_{1,2}$), FGs are believed to be ancient systems, that remained unperturbed for a long time. However, X-ray analyses of their intragroup gas show that they often lack the presence of a cool core (CC), which would argue against their hypothetical isolation and absence of mergers \citep[][ Dupke et al. in prep]{miller2012}\\

Given that FGs are principally found at $z<0.2$, recent efforts have been oriented to search for their progenitors to unveil their origin and mechanisms of formation. \cite{irwin2015} discovered that the Cheshire Cat cluster is indeed a fossil group progenitor (FGP). They also showed that the merger of two separate groups can originate a FG once the BCGs merge, which would explain the absence of CC in some of them. Simulations supported this alternative scenario too, placing many FGPs between $0.3<z<0.6$. Taking the Cheshire Cat as an archetype, the search for FGPs is driven by the following criteria \citep{johnson2018,kanagusuku2016}:
\begin{enumerate}
    \item Clusters/groups that lie in the interval $0.3<z<0.6$.
    \item Strong lenses, since FGs usually display large concentrations of matter thanks to their old ages \citep{wechsler2002}.
    \item Clusters/groups with an ongoing or imminent major merging that could give birth to a massive FG BCG within the system's lookback time.
\end{enumerate}

In the case of WHL0137, we inferred a richness-size ($N-R_{200}$) scaling relation using the spectra collected for the CLASH clusters and described in Sect. \ref{spec_members}. Applied to WHL0137, it yields a virial radius of $R_{200}=1.08$ Mpc, which is compatible with the value of $R_{200}=1.24$ Mpc estimated by \cite{wen2015} using a luminosity-size scaling relation calculated from SDSS data. Using the F606W ACS/WFC filter as a proxy for the $r$-band, we found magnitude gaps of $\Delta m_{1,2}=1.69$ and $\Delta m_{1,4}=2.07$ within 0.5 $R_{200}$. As a consequence, WHL0137 does not satisfy the basic criteria to be classified as FG. However, as it fulfills the first two criteria to be a FGP and the original magnitude gaps are high, we estimated its potential to become a FG by comparing the merger timescales of the cluster members with the cluster's lookback time, as described by \cite{johnson2018}. In short, we determined which galaxy members will have time to merge with the BCG before z$\sim$0, and we reevaluated the magnitude gaps afterwards. Following the recipe developed by \cite{kitzbichler2008} for photometric redshifts, at WHL0137's redshift, the time to merge for a pair of galaxies can be computed as:
$$\langle T_{merge}\rangle=99.75\,r\,M_*^{-0.3}\,\,\mathrm{ Gyr}$$
\noindent where $r$ is the projected distance between the BCG an a certain galaxy member (in kpc) and $M_*$ is the sum of the galaxies' masses (in solar masses). We assumed a constant mass-to-light ratio for the F606W band of 6, in line with \cite{johnson2018}. \\

We found that 48 of the galaxy members that lie within 0.5$R_{200}$ will have time to merge with the BCG, increasing the brightness of the latter by 0.98 mag. As a result, the new magnitude gaps become $\Delta m_{1,2}=2.67$ and $\Delta m_{1,4}=3.13$. Thus, both Jones and Dariush criteria would be satisfied and WHL0137 is, tentatively, a FGP. We remark here that, in addition to the several approximations made, all these calculations are based on the (strong) assumption that no other luminous galaxy, with a difference in magnitude with the BCG smaller than 2 mag, will fall into the cluster during the lookback time. Additionally, it is also implicitly assumed that all the stars hosted by the galaxies that will potentially merge with the BCG will eventually end up in the BCG rather than becoming part of the ICL.\\

\section{Conclusions}\label{sect_conclusions}

We analysed the optical and NIR ICL fractions in WHL0137, a massive cluster of galaxies at $z=0.566$. We used the HST images observed by the RELICS project, given their superb quality. We disentangled the galactic light from the ICL using CICLE, a robust, precise, and free of a priori assumptions ICL mapper. The cluster membership was estimated from photometric properties with the machine learning algorithm GBM, taking CLASH photometry and spectra as training set given its resemblance to the RELICS data. From the ICL and total cluster light map we estimated the ICL fractions in the four optical F435/475/606/814W filters and the five IR F105/110/125/140/160W broad bands. Results can be summarized as follows:
\begin{itemize}
    \item ICL fractions for WHL0137 range between 6\% and 31\%. Filter F125W was rejected since it was not reliable due to its higher level of noise.
    \item ICL fractions in the visible ($\sim$6\%--19\%) are fully consistent with the merging regime found in \cite{jimenez-teja2018} for clusters in $0.18<z<0.55$ ($\sim$7\%--23\%). Moreover, the main signature of the dynamically active clusters, the excess in the ICL fraction measured in the F606W band, is also present. This excess is explained by invoking the tidal stripping of the stars in the outer layers of massive, infalling galaxies and the tidal disruption of dwarf galaxies during a merger event.
    \item ICL fractions in the IR are, on average, $\sim$50\% higher than the optical ones (excluding F435W), which indicates that the ICL is dominated by an old and/or higher-metallicity population, probably drawn from the older cluster members passively by dynamical friction, interactions with the BCG, intragroup light originated by preprocessing in infalling groups and/or the result of the aging of pre-existing, younger populations.
    \item Using the F606W HST/ACS filter as a proxy for the $r$-band, we calculate that WHL0137 is a potential fossil group progenitor.
\end{itemize}

This pilot study not only gave insights on the properties of the ICL of the cluster WHL0137 through the study of the ICL fraction, but also showed the potential of the RELICS data for ICL studies. A systematic, extensive, and unbiased analysis of the ICL fraction will be conducted using the whole sample of 41 clusters of RELICS.\\

\begin{acknowledgements}

This project has received funding from the European Union’s Horizon 2020 research and innovation programme under the Marie Skłodowska-Curie grant agreement No 898633. Y.J-T. and J.M.V. acknowledge financial support from the State Agency for Research of the Spanish MCIU through the
"Center of Excellence Severo Ochoa" award to the Instituto de Astrofísica de Andalucía (SEV-2017-0709). J.M.V. acknowledges support from project  PID2019-107408GB-C44 (Spanish Ministerio
de Ciencia e Innovaci\'on). R.A.D. acknowledges partial support support from NASA Grants 80NSSC20P0540  and 80NSSC20P0597 and the CNPq grant 308105/2018-4. P.A.A.L. thanks the support of CNPq, grant 309398/2018-5. N.O.L.O. acknowledges financial support from CNPq under the grant No. 141631/2020-1. This work is based on observations taken by the RELICS Treasury Program (GO 14096) with the NASA/ESA HST, which is operated by the Association of Universities for Research in Astronomy, Inc., under NASA contract NAS5-26555. 

\end{acknowledgements}

\bibliography{WHL0137.bib}{}
\bibliographystyle{aasjournal}

\end{document}